\documentclass[10pt, onecolumn,english,prd,showpacs,nofootinbib]{revtex4}
\usepackage[latin9]{inputenc}
\usepackage{graphicx}
\usepackage{amsmath}
\usepackage{amssymb}
\usepackage{amsfonts}
\usepackage{lscape}
\usepackage{url}
\usepackage{epsfig}
\usepackage{color}
\usepackage{bm}
\usepackage{babel}
\usepackage{slashbox}
\usepackage{xcolor}
\usepackage{tabularx}
\providecommand{\tabularnewline}{\\}


\newcommand{\nn}{\nonumber}
\newcommand{\newc}{\newcommand}
\newc{\be}{\begin{equation}}
\newc{\ee}{\end{equation}}
\newc{\ba}{\begin{eqnarray}}
\newc{\ea}{\end{eqnarray}}
\newc{\bea}{\begin{eqnarray*}}
\newc{\eea}{\end{eqnarray*}}
\def \lcdm {$\Lambda$CDM }
\def \lcdmb {$\Lambda$CDM}
\def \om {\Omega_{\rm m}}
\def \omo {\Omega_{\rm m_0}}
\def \rmd {{\rm d}}
\def \hti {\tilde{h}}

\begin{document}
\title{Comparison of piecewise-constant methods for dark energy}
%\title{Comparison of Principal Component Analysis methods for Dark Energy}
%\title{Magic mirror on the wall, who is the best PCA of them all?}

\author{Savvas Nesseris}
\email{savvas.nesseris@uam.es}

\author{Domenico Sapone}
\email{domenico.sapone@uam.es}

\affiliation{Departamento de F\'isica Te\'orica and Instituto de F\'isica Te\'orica, \\
Universidad Aut\'onoma de Madrid IFT-UAM/CSIC,\\ $28049$ Cantoblanco, Madrid, Spain}


\begin{abstract}
We compare four different methods that can be used to analyze the type Ia supernovae (SnIa) data, ie to use piecewise-constant functions in terms of: the dark energy equation of state $w(z)$, the deceleration parameter $q(z)$, the Hubble parameter $H(z)$ and finally the luminosity distance $d_L$. These four quantities cover all aspects of the accelerating Universe, ie the phenomenological properties of dark energy, the expansion rate (first and second derivatives) of the Universe and the observations themselves. For the first two cases we also perform principal component analysis (PCA) so as to decorrelate the parameters, while for the last two cases we use novel analytic expressions to find the best-fit parameters. In order to test the methods we create mock SnIa data (2000 points, uniform in redshift $z\in[0,1.5]$) for three fiducial cosmologies: the cosmological constant model ($\Lambda$CDM), a linear expansion of the dark energy equation of state parameter $w(a)=w_0+w_a(1-a)$ and the Hu-Sawicki $f(R)$ model. We find that if we focus on the two mainstream approaches for the PCA, i.e. $w(z)$ and $q(z)$, then the best piecewise-constant scheme is always $w(z)$. Finally, to our knowledge the piecewise-constant method for $H(z)$ is new in the literature, while for the rest three methods we present several new analytic expressions.
\end{abstract}
\maketitle

\section{Introduction}

We live in an epoch at which several theories have been built to explain the observed accelerated expansion of the Universe, see Ref.~\cite{sapone_review} for a review. The discovery of this late-time acceleration of the Universe by \cite{sn1a} has led people to introduce a new ingredient to total matter density in the Universe: the dark energy. Such a component has raised severe, and still unsolved, theoretical problems which have led the community to search for alternative approaches to explain this late-time acceleration, namely modified gravity models and inhomogeneous models. The first approach simply postulates that general relativity is accurate up to a typical scale and it needs to be modified at larger scales; this modification could then lead to an observed acceleration. The second approach studies the effects of the large scale structures on the observed luminosity of distant supernovae of type Ia (SnIa).

Huge experimental efforts have been made to better understand the expansion of the Universe, see for instance \cite{Cimatti:2009is},\cite{euclid-red} \cite{AbbottBI},
\cite{PAU1}, \cite{PAU2} for details on the surveys of Euclid, DES and PAU. All of these experiments are planned to collect an impressive amount of data for different observables to reduce the statistical errors on the cosmological parameters. Once the data from the different experiments are collected, then we need to be able to extract the largest amount of information on the cosmological parameters by using the smallest number of assumptions possible.

This paper looks towards this direction: we use the principal component analysis (PCA) in order to decorrelate the parameters of interest and get unbiased constraints. Another advantage of the PCA is that there is no need to specify the cosmology; i.e. it is a model-independent approach, as the PCA will give us a set of functions that better describes the data \cite{Huterer:2002hy}, \cite{dePutter:2007kf}, \cite{Mortonson:2009hk}, \cite{Hojjati:2011xd},  \cite{hutererch}, \cite{Sapone:2014nna}. More details on the procedure of the PCA and the relevant sets of equations for each approach are also presented in Sec. \ref{sec:background} and in the Appendix.

In this paper in particular, we will compare different forms of the PCA and analyze their advantages and disadvantages. The four different cases we will consider are
\begin{enumerate}
  \item the deceleration parameter $q(z)$,
  \item the DE equation of state (EOS) parameter $w(z)$,
  \item the Hubble parameter $H(z)$ and
  \item the luminosity distance $d_L(z)$.
\end{enumerate}
These quantities are the most phenomenologically interesting ones as they are connected directly with the physical properties of the DE fluid ($w(z)$), the expansion rate of the Universe ($H(z)$ and $q(z)$) and finally the measurements themselves ($d_L(z)$).

We then analyze the four mentioned parameterizations with three different mock SnIa data based on different cosmologies:
\begin{enumerate}%[I]
  \item The \lcdm model with $w=-1$;
  \item The $w_0\,w_a$CDM model with $\{w_0,w_a\}=\{-1.2,0.5\}$;
  \item The Hu and Sawicki $f(R)$~\cite{Hu:2007nk} and Nesseris et al \cite{Basilakos:2013nfa} model with $\{b,n\}=\{0.1,1\}$ parameters.
\end{enumerate}
For all the mock catalogs we assume that $\omo=0.3$ and $h=0.7$, while some more details about the construction of the mocks are given in Sec.~\ref{sec:results}.

As it will be quite clear in the next few sections, the novelty of our paper lies in the following two pillars: the systematic comparison of all the methods and their subsequent testing against mock data and against each other is done for the first time in the literature and second, we present several new analytical expressions for all four different forms of piecewise-constant methods. Our analysis will be immensely useful with the upcoming surveys that will collect a plethora of new data that will have to be analyzed in a systematic fashion and their cosmological information extracted.

Finally, the paper is organized as follows: in Sec. \ref{sec:theory} we briefly review the background equations and the models we consider in this analysis, in Sec. \ref{sec:background} we briefly review the PCA and we present several novel results for all four forms of the PCA (see also the Appendix). In Sec.~\ref{sec:results} we present the results we found by applying the four different forms with the three mock catalogs; finally in Sec.~\ref{sec:conclusions} we summarize our conclusions, listing the advantages and disadvantages of all the forms of the PCA used in this paper.

\section{Theory}\label{sec:theory}
In this section we briefly review the equations for the \lcdmb, $w_0\,w_a$CDM and the Hu and Sawicki $f(R)$ models. The $w_0\,w_a$CDM model assumes a linear expansion in terms of the scale factor of the dark energy equation of state, such that
\be
w(a)=w_0+w_a(1-a).
\ee
The cosmological constant model \lcdm corresponds to $(w_0,w_a)=(-1,0)$. The Hubble parameter, assuming a flat universe, is given by
\be
H(z)^2/H_0^2=\omo (1+z)^3+(1-\omo)(1+z)^{3(1+w_0+w_a)}e^{-3 w_a z/(1+z)}.
\ee
The Hu and Sawicki $f(R)$~\cite{Hu:2007nk} is given by the action
\begin{equation}
S=\int d^{4}x\sqrt{-g}\left[  \frac{1}{2k^{2}}f\left(R\right)
+\mathcal{L}_{m}\right]  \label{action1}%
\end{equation}
where $\mathcal{L}_{m}$ is the Lagrangian of matter and $k^{2}=8\pi G$ and the function $f(R)$ is given by
\begin{equation}
\label{Hu}
f(R)=R-m^2 \frac{c_1 (R/m^2)^n}{1+c_2 (R/m^2)^n}\,.
\end{equation}
In Nesseris et al \cite{Basilakos:2013nfa} it was shown that this can also be rewritten as \ba
\label{Hu1}
f(R)&=& R- \frac{m^2 c_1}{c_2}+\frac{m^2 c_1/c_2}{1+c_2 (R/m^2)^n} \nn\\
&=& R- 2\Lambda\left(1-\frac{1}{1+(R/(b~\Lambda)^n}\right) \nn \\
&=& R- \frac{2\Lambda }{1+\left(\frac{b \Lambda }{R}\right)^n}
\ea where $\Lambda= \frac{m^2 c_1}{2c_2}$ and $b=\frac{2 c_2^{1-1/n}}{c_1}$. In this form it is obvious that the Hu and Sawicki model can be arbitrarily close to $\Lambda$CDM, depending on the parameters $n$ and $b$. The explicit modified Friedmann equations can be found in Ref. \cite{Basilakos:2013nfa}.

\section{The PCA}\label{sec:background}

Let us suppose that we have a function $f(x)$ well defined in the range $x_a$ and $x_b$
and we want to find the best parametrization to this function given a set of data $D_i$.
We can write the function $f(x)$ with many piecewise-constant values as:
\be
f(x) = \sum_{i=i}^{N}f_i\theta_i(x)
\label{eq:piecewise}
\ee
where $f_i$ are constant in each interval $x_i$ and $\theta(x_i)$ is the theta function, i.e.
$\theta(x_i) = 1$ for $x_{i-1}< x \leq x_{i}$ and $0$ elsewhere. $N\gg 1$ is the number of parameters
that we would like to constrain given the data. However, the parameters will be in general correlated; we can use the PCA to decorrelate the parameters $f$'s. Following Ref.~\cite{hutererch}, we first build a diagonal matrix $\Lambda_{ij}$ with the eigenvalues of the Fisher matrix $F_{ij}$, which is defined as the inverse of the covariance matrix $C_{ij}$ (obtained directly from the chains, when performed).
Then we define a matrix $\tilde{W}_{ij}=W_{ik}^{T}\,\Lambda_{km}^{1/2}\,W_{mj}$ where
the matrix $W_{km}^{T}$ is the transpose of $W_{km}$ and the latter is a matrix
composed by the eigenvectors of Fisher matrix. We finally normalize $\tilde{W}_{ij}$
such that its rows sum up to unity. The matrix $\tilde{W}_{ij}$ will give the
uncorrelated parameters, i.e.
\be
p_i=\sum_{j=1}^{M}\tilde{W}_{ij}\,f_{j}
\label{eq:pca-values}
\ee
where $M$ isthe total number of parameters. The variance of the parameters $p_i$ will then be
\be
\sigma^2\left(p_i\right)=1/\lambda_i\,.
\label{eq:pca-errors}
\ee
Our goal is to investigate which is the best parametrization that gives us the
largest amount of information about the cosmology given a set of observations. So, in what follows we consider different parameterizations, such as: the deceleration parameter $q(z)$, the dark energy EOS $w(z)$, the Hubble parameter $H(z)$ and the luminosity distance $D_L(z)$. We find the best fit and consequently the principal components (PC) of each parametrization and we then propagate this results to the other parameters. As an example, we find the best fit and the PC for the deceleration parameter $q$ and then we convert the results to all the other parameters $w$, $H$ and $D_L$. In this way we are able to verify which is the optimal parametrization with which we can gain most of the information hidden in the data.

To clarify the analysis, let us consider a function $\psi(z)$, where $z$ is the redshift, and it can be used to calculate the luminosity distance $d_L(z) = d_L(\psi(z))$. As a first step we assume that the function $\psi(z)$ can be approximated as piecewise-constant in redshift bins, as in Eq.~(\ref{eq:piecewise}); then we evaluate the luminosity distance $d_L$ which now will be a function of the constant $\psi_n$:
\be
d_L(z) = d_L(z; \psi_1, \psi_2,...,  \psi_n)\,.
\ee
In this case the values of the $d_L$ at one redshift will depend on all the $\psi$'s in the previous redshifts, hence each value of $\psi_i$ will appear in several bins and consequently the $\psi$'s will be correlated between different bins.
The reason why we used the PCA is to decorrelate these parameters and to extract the maximal amount of information for the parameters. The analysis is based on the algebraic concept to find a linear transformation that it is able to diagonalize the covariance matrix.

\subsection{The deceleration parameter $q(z)$\label{PCAq}}

We start by presenting the results of the piecewise-constant deceleration parameter $q$, derived in Ref.~\cite{nesserisBIA}:
\be
q(z) = \sum_{i=1}^{n}q_i\theta(z_i),
\ee
where $q_i$ are constant in each redshift bin $z_i$ and $\theta(z_i)$ is the theta function defined before. Once the PC of the $q_i$ parameters are found we can derive the other observables using the definition of the deceleration parameter:
\be
1+q(z)=\frac{d \ln(H(z))}{d\ln(1+z)}\,.
\label{eq:q}
\ee
The Hubble parameter and the luminosity distance are (see Appendix \ref{app-PCAq} for more details):
\ba
H_n(z) &=& H_0 b_n \left(1+z\right)^{1+q_n}\label{eq:hubble-q} \\
d_{L,n}(z)&=& \frac{c}{H_0}\left(1+z\right)\left[f_n-\frac{\left(1+z\right)^{-q_n}}{b_n q_n}\right]
\label{eq:dl-q}
\ea
where the coefficients $b_n$ and $f_n$ are
\ba
b_n &=& \prod_{j=1}^{n-1}\left(1+z_j\right)^{q_j-q_{j+1}}\\
f_n &=&  \frac{\left(1+z_{n-1}\right)^{-q_n}}{b_n q_n}+\sum_{j=1}^{n-1}\frac{\left(1+z_{j-1}\right)^{-q_j}-
\left(1+z_{j}\right)^{-q_j}}{b_j q_j}
\ea
and $z_0=0$.

To propagate our results into the dark energy EOS parameter $w$,
we make use of Eq~(\ref{eq:q}) and we express the Hubble parameter as
\be
E^2(z) = H^2(z)/H^2_0 = \omo (1+z)^3 + \left(1-\omo\right)e^{3\int_{0}^{z}\frac{1+w(x)}{1+x}\rmd x}\,.
\label{eq:hubble-gen}
\ee
Deriving the last equation we then find
\be
w_n(z) = \frac{1}{3}\frac{2q_n-1}{1-\om(z)}
\label{eq:w-fromq}
\ee
where $\om (z)$ is the matter density as a function of redshift:
\be
\om(z) = \frac{\omo (1+z)^3}{H_n^2(z)/H_0^2}\,.
\label{eq:omega_m}
\ee
It is important to notice that Eq.~(\ref{eq:w-fromq}) depends on the matter density $\omo$, for which we have to assume a specific value not derived from the data themselves (we will come back later on this).

\subsection{The dark energy equation of state $w(z)$\label{PCAw}}

We now want to apply the PCA directly to the dark energy EOS parameter $w(z)$. As previously done for $q$, we rewrite $w(z)$ as
\be
w(z) = \sum_{i=1}^{n_{tot}}w_i\theta(z_i),
\ee
where $w_i$ are constant in each redshift bin.
Using the energy-momentum conservation $\nabla_\mu T^{\mu\nu}=0$ for an ideal fluid with
equation of state $w$, we get the equation for the DE density can then be written,
for $z$ in the nth bin, as
\be
\rho_{DE}(z,n)=\rho_{DE}(z=0) c_n \left(1+z\right)^{3(1+w_n)},
\ee
where the coefficient $c_n$ is
\be
c_n = \prod_{j=1}^{n-1}\left(1+z_j\right)^{w_j-w_{j+1}}\,.
\ee
Consequently, we can write the Hubble parameter and the luminosity distance as
(see Appendix \ref{app-PCAw} for more details)
\ba
H_n(z)^2/H_0^2 &=& \omo (1+z)^3+(1-\omo)c_n \left(1+z\right)^{3(1+w_n)} \label{eq:hubble-w} \\
d_{L,n}(z) &=& \frac{c}{H_0} (1+z)\left(d_n(z,z_{n-1})+\sum_{i=1}^{n-1}d_i(z_i,z_{i-1})\right)\,,
\label{eq:dl-w}
\ea
where
\ba
&&d_i(z_i,z_{i-1})\equiv\int_{z_{i-1}}^{z_i}\frac{\rmd z}{\sqrt{\omo (1+z)^3+(1-\omo)c_i \left(1+z\right)^{3(1+w_i)}}}=\nn\\
&& -\frac{2}{\omo^{1/2}} \left\{ \frac{ _2F_1\left[\frac{1}{2}, -\frac{1}{6w_i}, 1-\frac{1}{6w_i};
-c_i\frac{1-\omo}{\Omega_{m_0}}\left(1+z_i\right)^{3w_i}\right]}{\sqrt{1+z_i}}
 - \frac{ _2F_1\left[\frac{1}{2}, -\frac{1}{6w_i}, 1-\frac{1}{6w_i};
-c_i\frac{1-\omo}{\omo}\left(1+z_{i-1}\right)^{3w_i}\right]}{\sqrt{1+z_{i-1}}} \right\} ~~
\ea
and ${}_2F_1(a,b;c;z)$ is a hypergeometric function \cite{handbook}. Then, the deceleration parameter $q$ will be given by inverting Eq.~(\ref{eq:w-fromq}), i.e.
\be
q(z)=\frac12+\frac32 w\bigg[1-\om (z)\bigg]
\label{qfromw}
\ee
where $\om(z)$ is given by Eq.~(\ref{eq:omega_m}) for which we use the Hubble parameter in Eq.~(\ref{eq:hubble-w}).

\subsection{The Hubble parameter $H(z)$}

In order to apply the PCA to the Hubble parameter $H(z)$, we write it as
\be
H_n(z)/H_0 = \sum_{i=1}^{n_{tot}}h_i\theta(z_i),
\ee
where $h_i$ are constant in each redshift bin.
Using the definition of the luminosity distance along with the previous equations we have
\be
d_{L,n}(z)=\frac{c}{H_0} (1+z) \left(g_n+h_n^{-1} z\right),
\label{eq:ref-dl-h}
\ee
where we have defined the constants $g_n\equiv\sum_{i=1}^{n-1}z_i(h_i^{-1}-h_{i+1}^{-1})$.

In order to find the dark energy EOS $w$ we can use Eq.~(\ref{eq:hubble-gen}) and we find
\ba
w_n(x_i)&=&-1+\frac{\ln\left(\frac{H_i^2/H_0^2-\omo (1+z_{eff,i})^3}{H_{i-1}^2/H_0^2-
\omo (1+z_{eff,i-1})^3}\right)}{\ln\left(\frac{1+z_{eff,i}}{1+z_{eff,i-1}}\right)^3}, \nn \\
x_i &\simeq& z_{i-1} \label{wzeff}
\ea
where $z_{eff}$ is the average redshift of the bin, i.e. $z_{eff}\simeq (z_{i-1}+z_i)/2$
(see Appendix \ref{app-PCAh} for more details). Also in this case the dark energy EOS parameter depends on the matter density $\omo$ and we need to fix it to a particular value.

To obtain the deceleration parameter $q$ from the measurements of Hubble parameters $h_n$, we start:
\be
\frac{1+q(z)}{1+z} = \frac{H'(z)}{H(z)}\,,
\label{eq:q-H-func}
\ee
which can be integrated in the $j$th bin by assuming $q_j$ constant in that bin and we find:
\be
\left(1+q_j\right)\int_{z_{eff,j-1}}^{z_{eff,j}}{\frac{1}{1+z}{\rmd z}} = \int_{H_{j-1}}^{H_{j}}{\frac{1}{H}{\rmd H}}
\label{eq:q-from-h-0}
\ee
where $H_j$ is the Hubble parameter evaluated at the $j$ bin.
So we find the deceleration parameter to be:
\be
q_j = -1+\frac{\ln\left[H_j/H_{j-1} \right]}{
\ln\left[\left(1+z_{eff,j}\right)/\left(1+z_{eff,j-1}\right) \right]}\,.
\label{eq:q-from-h-1}
\ee
Similarly with before, see also the Appendix \ref{app-PCAh} for more details, we note that the parameters $H_j$ are evaluated in the effective redshift $z_{eff}$, but the resulting parameters are evaluated at the sides of the bins $x_i \simeq z_{i-1}$.
It is also interesting to notice that we can also evaluate the deceleration parameter by applying the definition of the derivative
to the Eq.~(\ref{eq:q-H-func}); the deceleration parameter becomes simpler and we can also avoid the problem of
defining the mean redshift:
\be
q_j = -1 + \frac{1 + z_j}{dz} \left(\frac{H_{j + 1}}{H_j} - 1\right)
\label{eq:q-from-h-2}
\ee
being $dz$ the bin width. We checked both Eqs.~(\ref{eq:q-from-h-0}) and (\ref{eq:q-from-h-1}) and we found that the results are almost identical. However, in this work we prefer to use Eq.~(\ref{eq:q-from-h-0}) to evaluate the deceleration parameter as
the definition of the derivative applies only when the infinitesimal quantity $dz$ in Eq.~(\ref{eq:q-from-h-1}) is small enough.

\subsection{The luminosity distance $d_L(z)$}

Let us write the luminosity distance $d_L(z)$ as
\be
d_{L,n}(z) = \sum_{i=1}^{n_{tot}}d_{L,i}\theta(z_i),
\ee
where $d_{L,i}$ are constant in each redshift bin.

Similarly with before, see also the Appendix \ref{app-PCAdl} for more details, we note that the parameters $d_{L,i}$ are evaluated in the effective redshift $z_{eff,i}$. Therefore, in order to get an estimate for $w(z)$ and $q(z)$ in this case, we can follow the same procedure as before. Using the definition of the luminosity distance at two redshifts $z_{eff,i}$ and $z_{eff,i-1}$ we have
\ba
\frac{d_{L,i}}{1+z_{eff,i}}-\frac{d_{L,i-1}}{1+z_{eff,i-1}}&=&\int_{z_{eff,i-1}}^{z_{eff,i}}\frac{c}{H(x)} \rmd x= \frac{c}{H(x)} \left(z_{eff,i}-z_{eff,i-1}\right),  \nn\\
x_i &\simeq& z_{i-1}
\ea
from which we can estimate $H(x)$ at the $i$th bin and finally estimate $w_i$ using Eq.~(\ref{wzeff}). In this case the values $w_i$ will correspond to the $z_{eff}$ redshift and not $z_i$. We have successfully tested these results with numeric tests and as before, we assume that our models do not have any fast transitions.

However, in this case we cannot discriminate between a constant $d_L(z)$ and a constant distance modulus $\mu(z)$, since these two are connected via
\be
\mu_{th}(z)= 5 \log_{10} D_L(z)+\mu_0
\label{distmod}
\ee
where $D_L(z)$ the dimensionless luminosity distance and
$\mu_0=5\log_{10}(\frac{\frac{c}{H_0}}{\textrm{Mpc}})+25\simeq42.384-5\log_{10}h$.
For the same reason we cannot differentiate $d_{L,i}$ from $\mu_0$,
since the latter is just a rescaling of the normalization, unless we fix $h$.
For these reasons, in what follows we will consider piecewise-constant $\mu$
and by assuming a value for $h$ we can later convert the best fits to $d_L$ and
use the previous relations to extract the cosmology. So, let us write the distance modulus $\mu(z)$ as
\be
\mu(z) = \sum_{i=1}^{n_{tot}}\mu_{i}\theta(z_i),
\ee
where $\mu_i$ are constant in each redshift bin (see Appendix~\ref{app-PCAdl} for more details). By making the distance modulus $\mu(z)$ piecewise-constant we have a diagonal covariance matrix which means that the parameters are already uncorrelated and we do not have to follow the PCA approach in this case (as the advantage to use the PCA is to decorrelate the parameters, i.e. make the covariance matrix diagonal).

In order to extract the cosmology we can invert Eq.~(\ref{distmod}) to find the dimensionless luminosity distance as
\be
D_{L,i}=10^{\frac{\mu_i-\mu_0}{5}}.
\label{eq:ref-dl-mu}
\ee
where $\mu_0\simeq42.384-5\log_{10}h$. We show these results in the next section.

\begin{figure}
\includegraphics[scale=0.7]{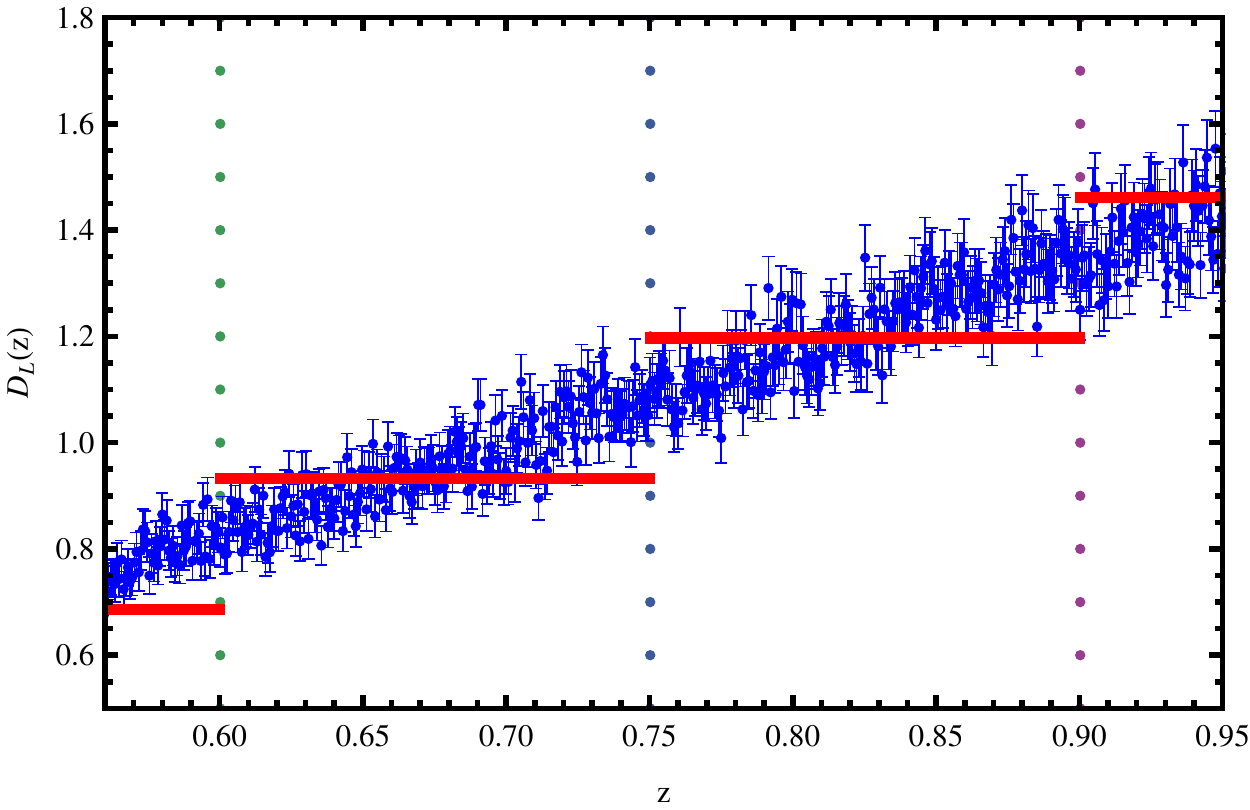}
\includegraphics[scale=0.7]{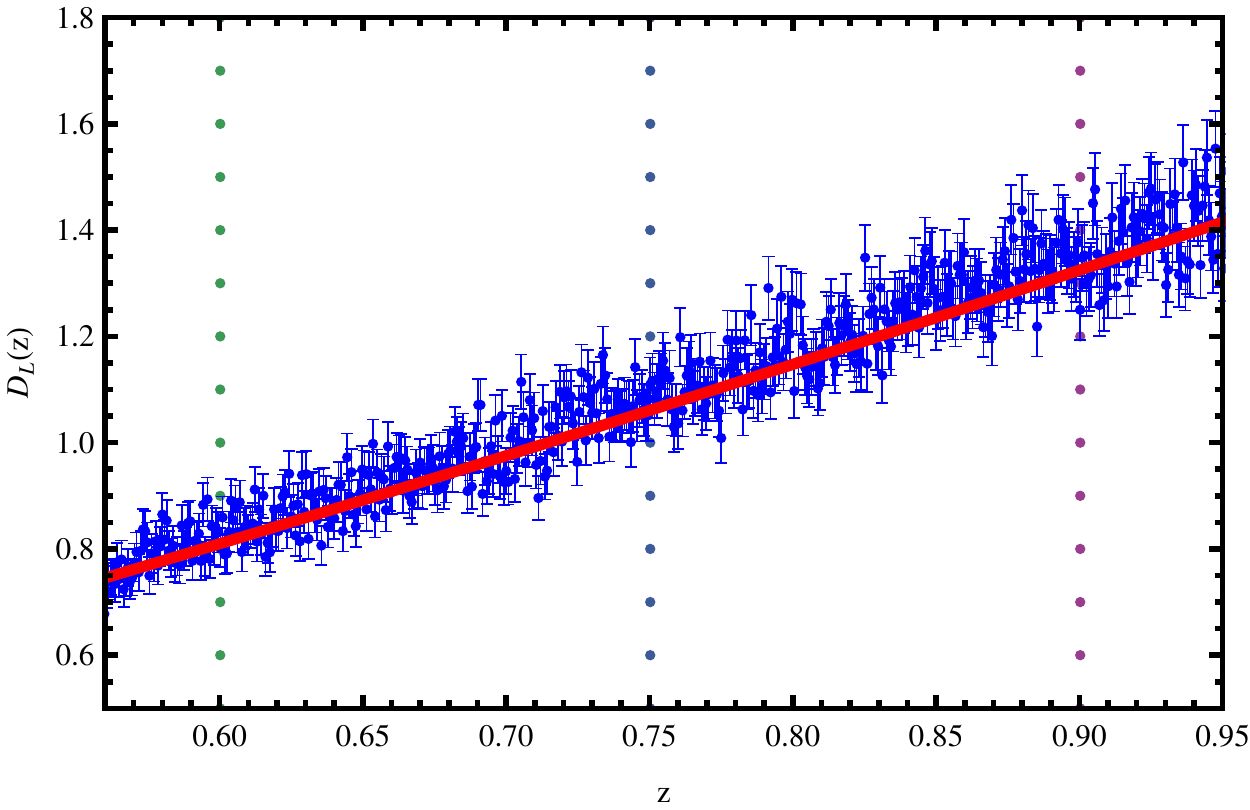}
\caption{The luminosity distance when $D_L$ is piecewise-constant in each bin (left) and when $q$ is piecewise-constant in each bin. In the first case, luminosity distance clearly has a discontinuity at the edge of the bins, while in the second case it is continuous. This discontinuity causes the large $\chi^2$ as mentioned in Table II. }
\label{fig:dlvsdl}
\end{figure}

However, we should mention that since $D_L$ is discontinuous on the redshift shell boundaries, this will lead to a large $\chi^2$ because of the assumptions, i.e. at the shell edges the model $D_L$ is bound to have a $\chi^2$ which becomes too large if there is enough data. This effect is illustrated in Fig. \ref{fig:dlvsdl}, where we compare the luminosity distance when $D_L$ is piecewise-constant in each bin (left) and when $q$ is piecewise-constant in each bin. In the first case, the luminosity distance clearly has a discontinuity at the edge of the bins, while in the second case it is continuous. As can be seen in the plot, the difference between the best-fit value (horizontal red line) and the data points near the edge of the shells is quite big, thus leading to a large $\chi^2$.

However, despite of this limitation we decided to also include this parametrization in our analysis for several reasons. First, as can be seen in Appendix \ref{app-PCAdl} this parametrization is equivalent to directly binning the SnIa. Second, our goal is to use as many different parametrizations as possible in order to cover all phenomenological properties of dark energy, going from the very fundamental $w(z)$, to the expansion of the Universe ($H(z)$ and $w(z)$) to the observations themselves ($D_L(z)$).

\section{Results}\label{sec:results}

In order to compare the different methods we created mock SnIa data based on an \textit{a priori} known cosmologies corresponding to the \lcdm, $w_a$CDM and $f(R)$ models. Since we are more interested in testing the methods themselves rather than worrying if the differences are due to the construction of the data, we evaluated $2000$ distance moduli uniformly distributed in the range $z\in[0,1.5]$; the distance modulus $\mu_{th}(z)$ was estimated as its theoretical value plus
a gaussian error (that can be negative or positive) and constant errors of $0.1$. Also, we should stress that since this is the first time this comparison appears in the literature we  implement the simplest possible way to produce mock data, as we are mainly concerned with comparing the different methods and not eliminating all possible sources of error in the data. Therefore, using more realistic mock SnIa data but also other kinds of data has been left for future work, since as mentioned before our current focus is the comparison of all the methods.

As mentioned previously, in order to apply the PCA to the $q_n$, $w_n$, $H_n$ and $D_{L_n}$
we need to find first the best-fit, given a data set, for the parameters.
For our purpose, we divide the survey into 10 equally spaced redshift bins up to $z=1.5$. To determine the best fit parameters we proceed in two different ways:
\begin{itemize}
\item For the $q_n$ and $w_n$, we perform a Monte Carlo Markov chain (MCMC) method, implemented by using the code of \cite{nesserisweb}. In the analysis we used more than 50000 steps each for the parametrizations of $q(z)$ and $w(z)$.
\item For the $H_n$ and $D_{L_n}$, we simply evaluate the minimum of the $\chi^2$ analytically.
\end{itemize}
Therefore, we have four distinct piecewise-constant methods in order to fit the data: method 1 the deceleration parameter $q(z)$, method 2) the dark energy equation of state $w(z)$, method 3 the Hubble parameter $H(z)$ and method 4 the luminosity distance $D_L$.

The reason we treat the last two methods differently is that for the luminosity distance $D_L$, we do not have to perform the PCA as the covariance matrix is already diagonal,
i.e. the parameters are uncorrelated, while for the Hubble parameter $H(z)$ we found that
the PCA fails for two reasons: first, because the parameters $h_i^{-1}$ are highly
correlated in a manner that a linear transformation, i.e. the PCA, cannot disentangle them,
and also the Fisher matrix has a highly unusual structure with several elements repeated
across its rows and columns, see Eq.~(\ref{fisherH}) in the Appendix. Also, we found that due to the fact that the parameters $h_i^{-1}$ are highly correlated a MCMC approach also fails since the sampler was always stuck along the degenerate lines $\sim h_i^{-1}-h_{i-1}^{-1}$ and far away from the minimum.

Then, we use the best-fit values in each case and the formulas found in the Appendix for each method, in order to calculate the ``derived" parameters. For example, in the third method for the Hubble data, first we calculate the best fit and then we use the expressions to calculate the parameters $q$, $w$ and $D_L$. To summarize, our methodology is as follows:
\begin{enumerate}
  \item Find the best fits for all the methods, either with a MCMC or analytically.
  \item If needed, do the PCA to diagonalize the covariance matrix ($q$ and $w$ only).
  \item Find the derived parameters in each case.
  \item Compare the methods.
\end{enumerate}

\begin{figure}
\includegraphics[scale=1]{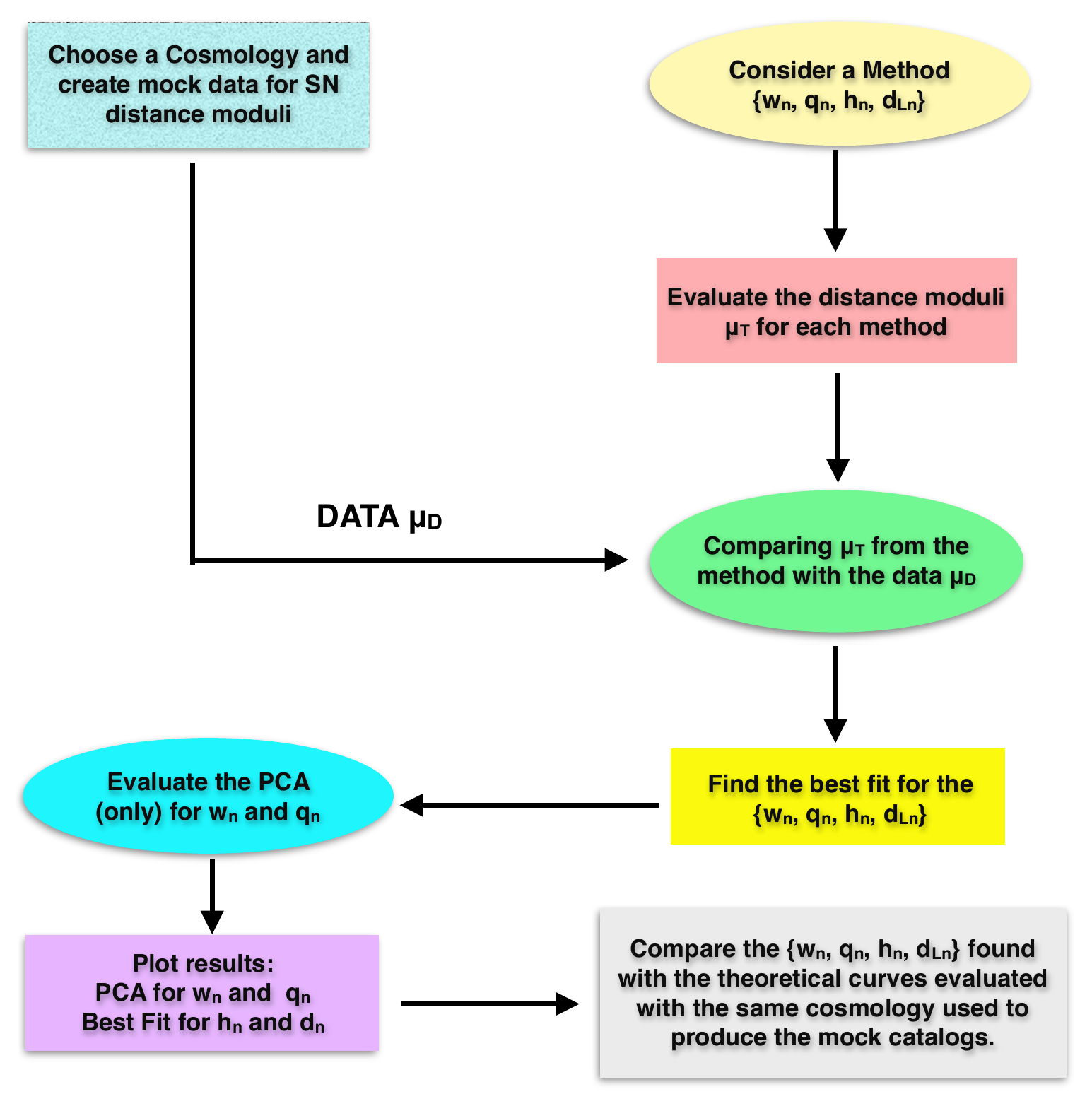}
\caption{A flowchart that shows the steps of the analysis in this paper.}
\label{fig:flowchart}
\end{figure}

\begin{figure}
\includegraphics[scale=0.45]{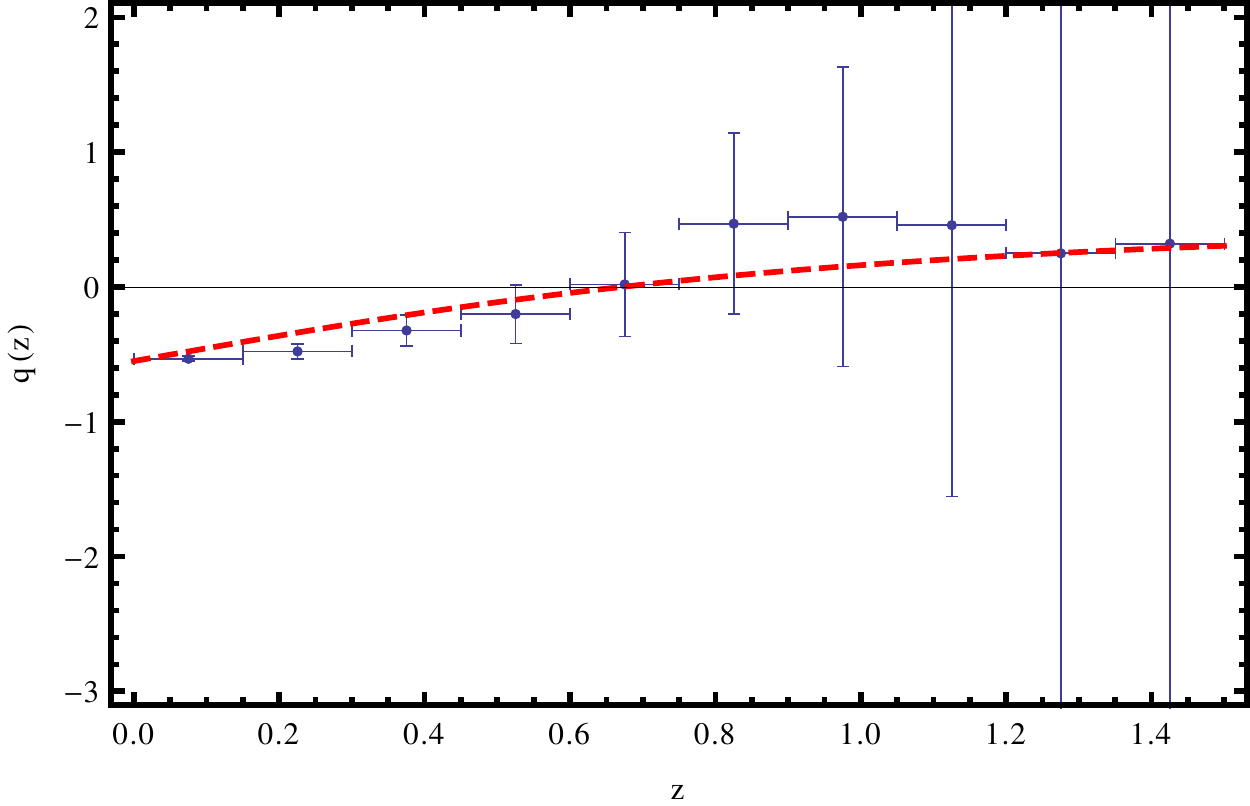}
\includegraphics[scale=0.45]{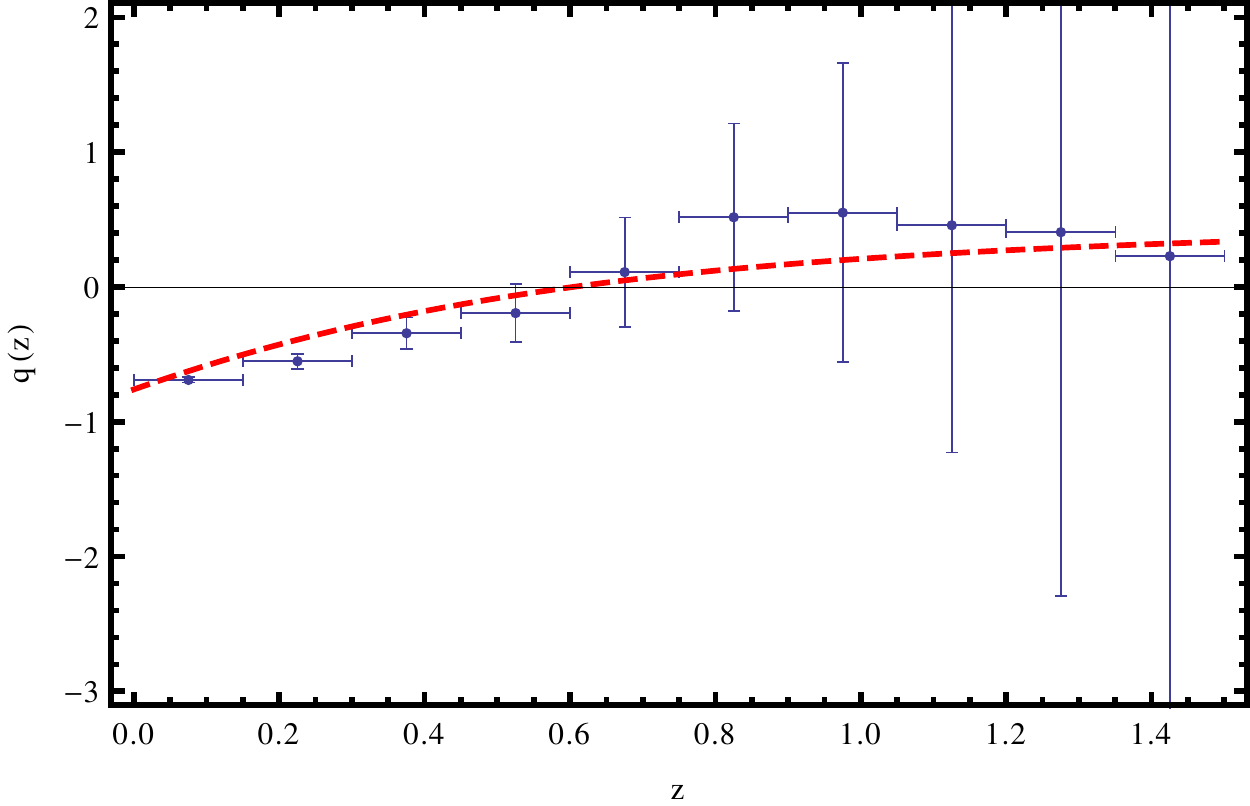}
\includegraphics[scale=0.45]{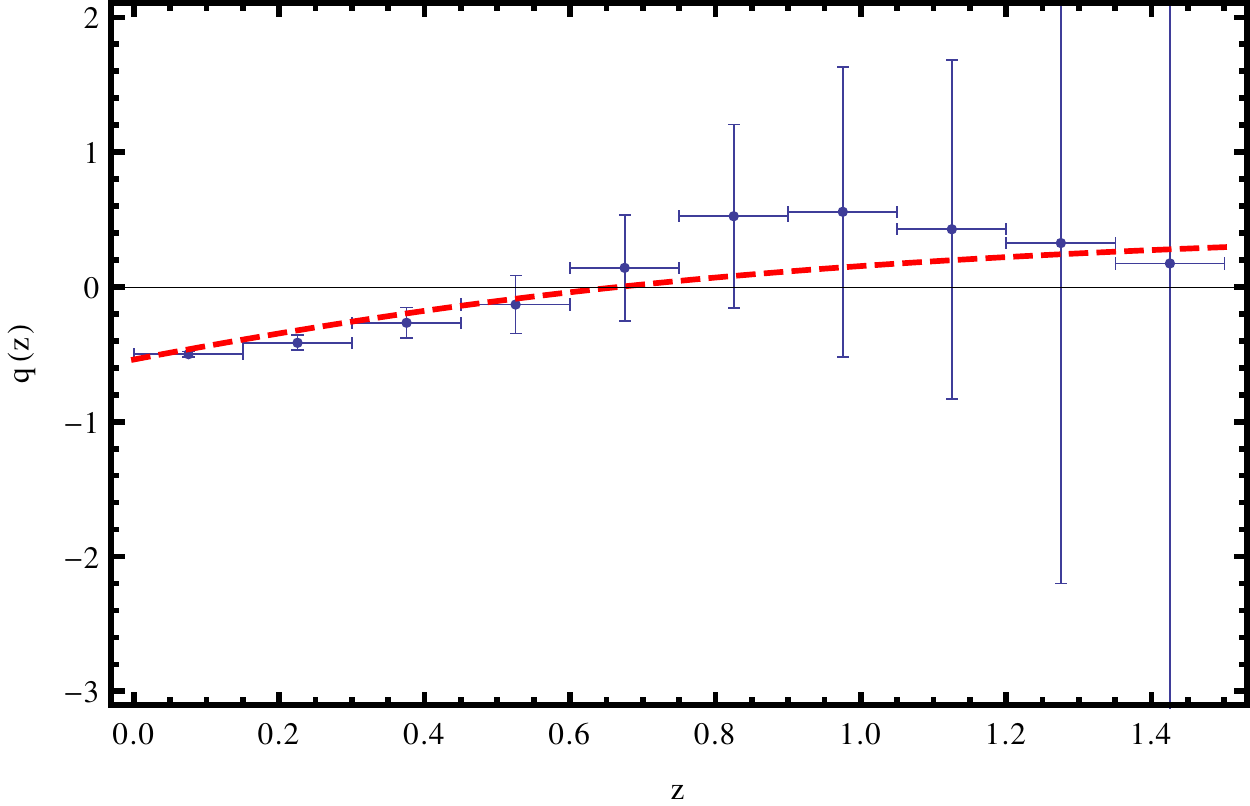}\\
\includegraphics[scale=0.45]{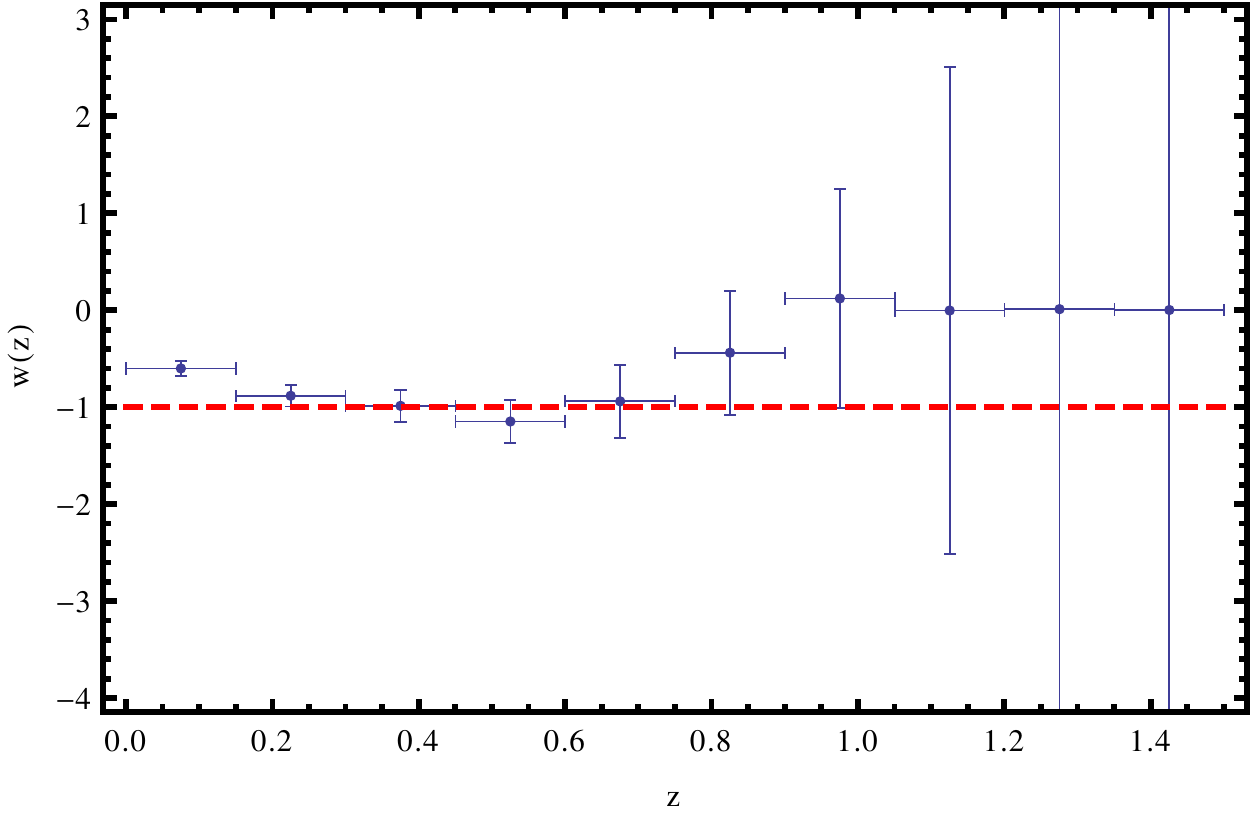}
\includegraphics[scale=0.45]{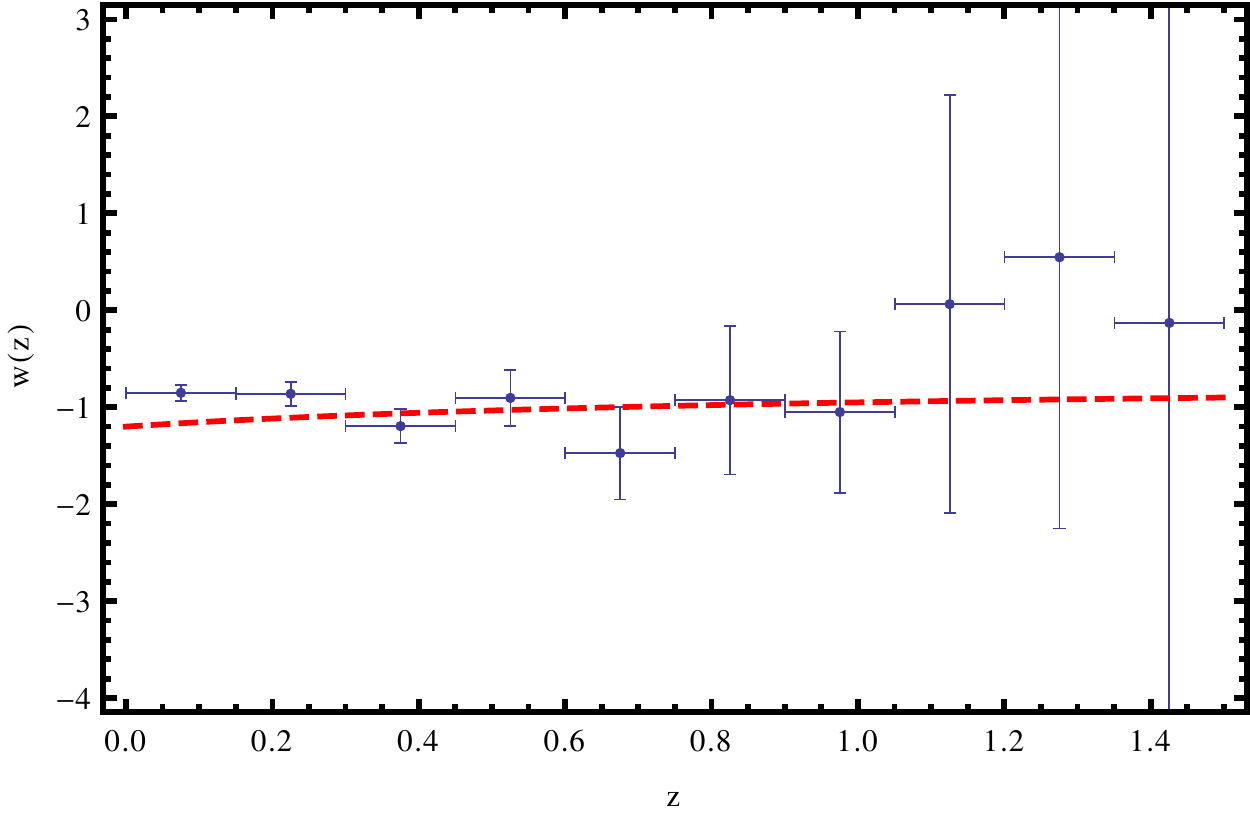}
\includegraphics[scale=0.45]{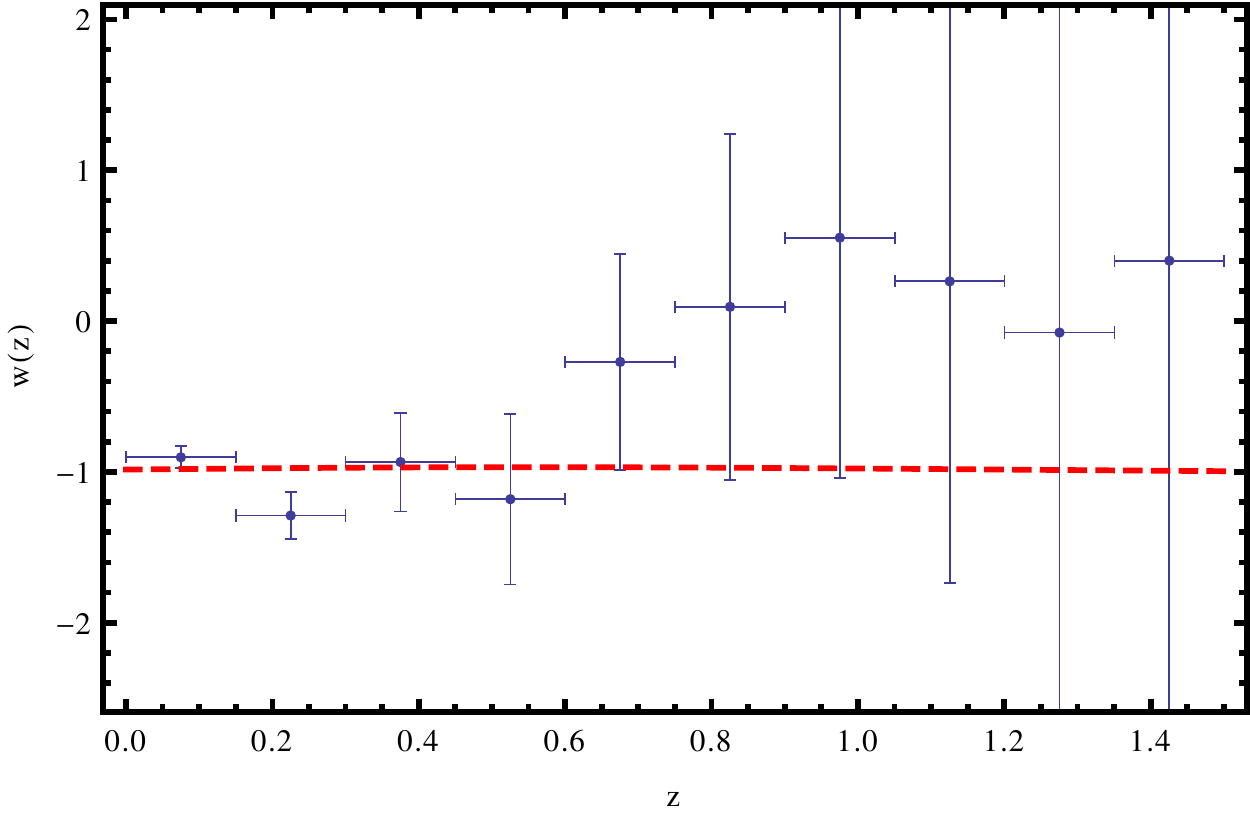}\\
\includegraphics[scale=0.45]{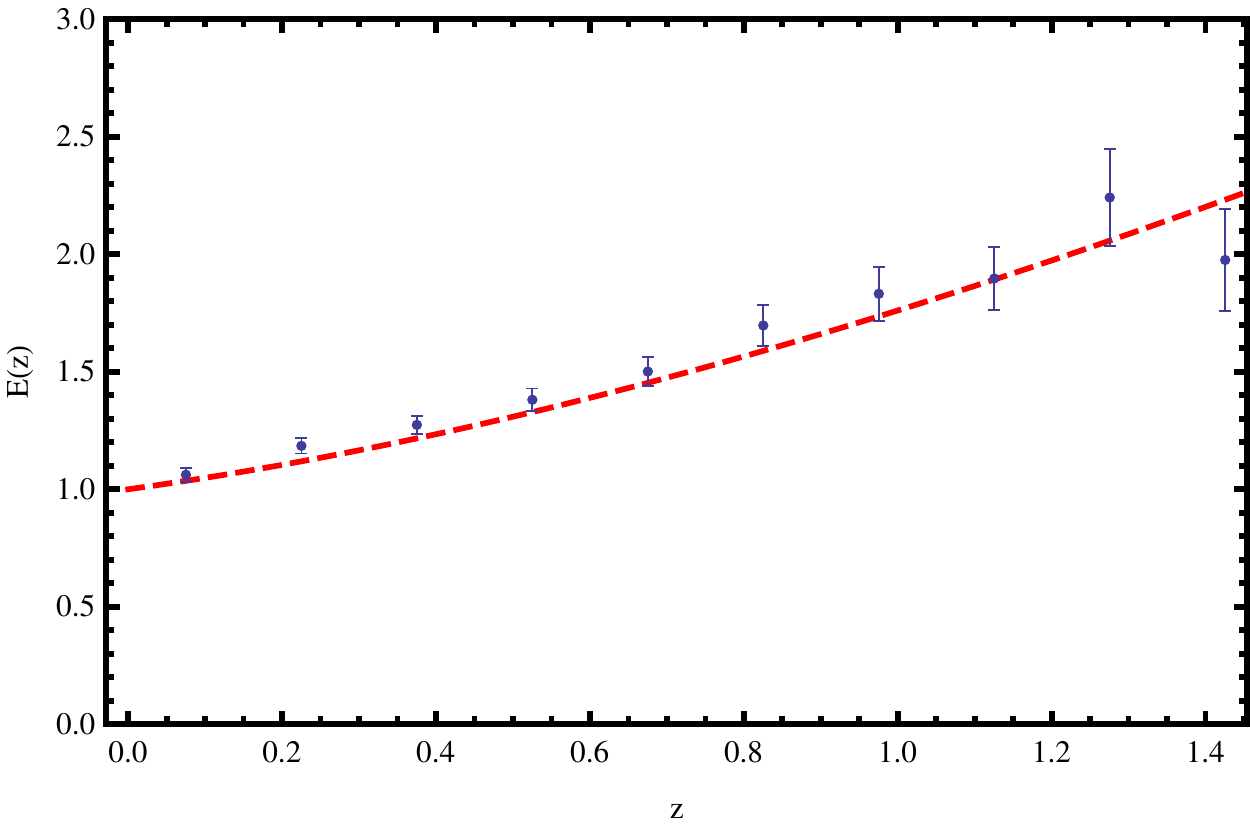}
\includegraphics[scale=0.45]{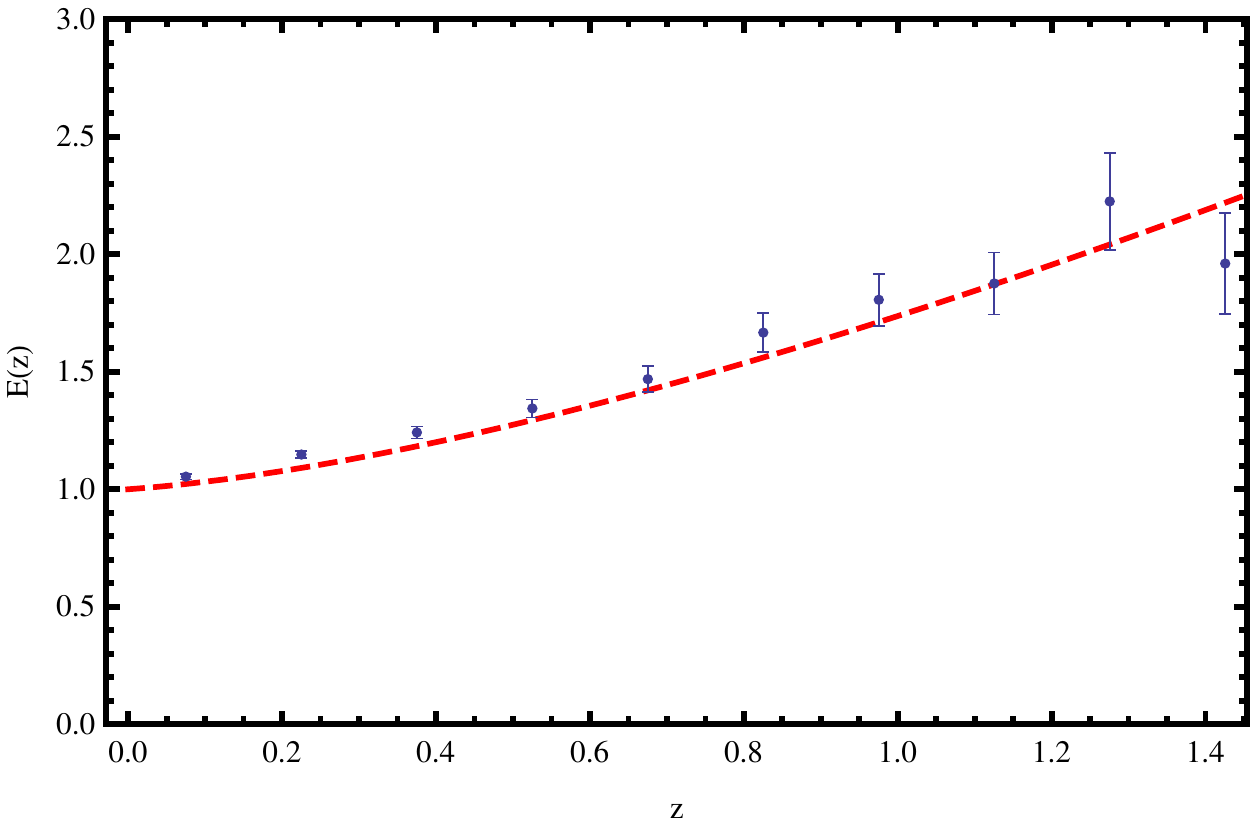}
\includegraphics[scale=0.45]{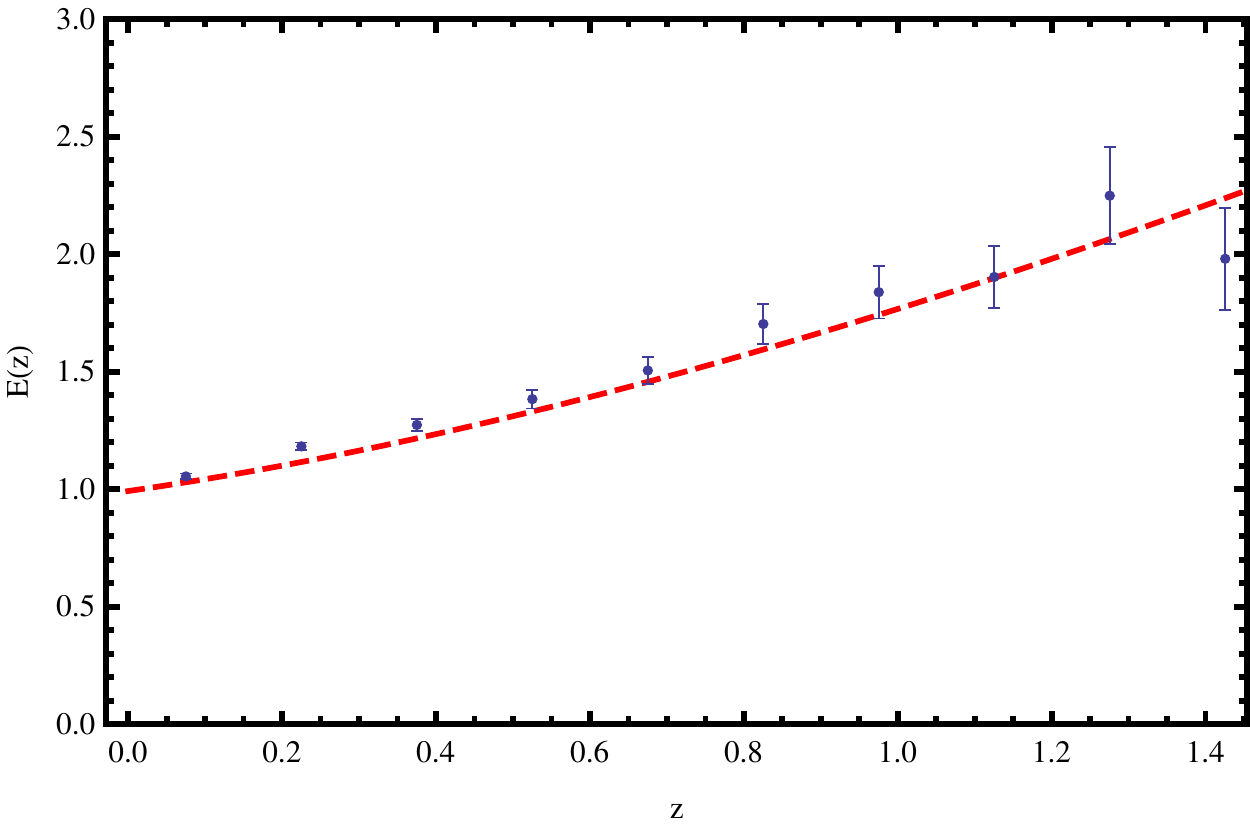}\\
\includegraphics[scale=0.45]{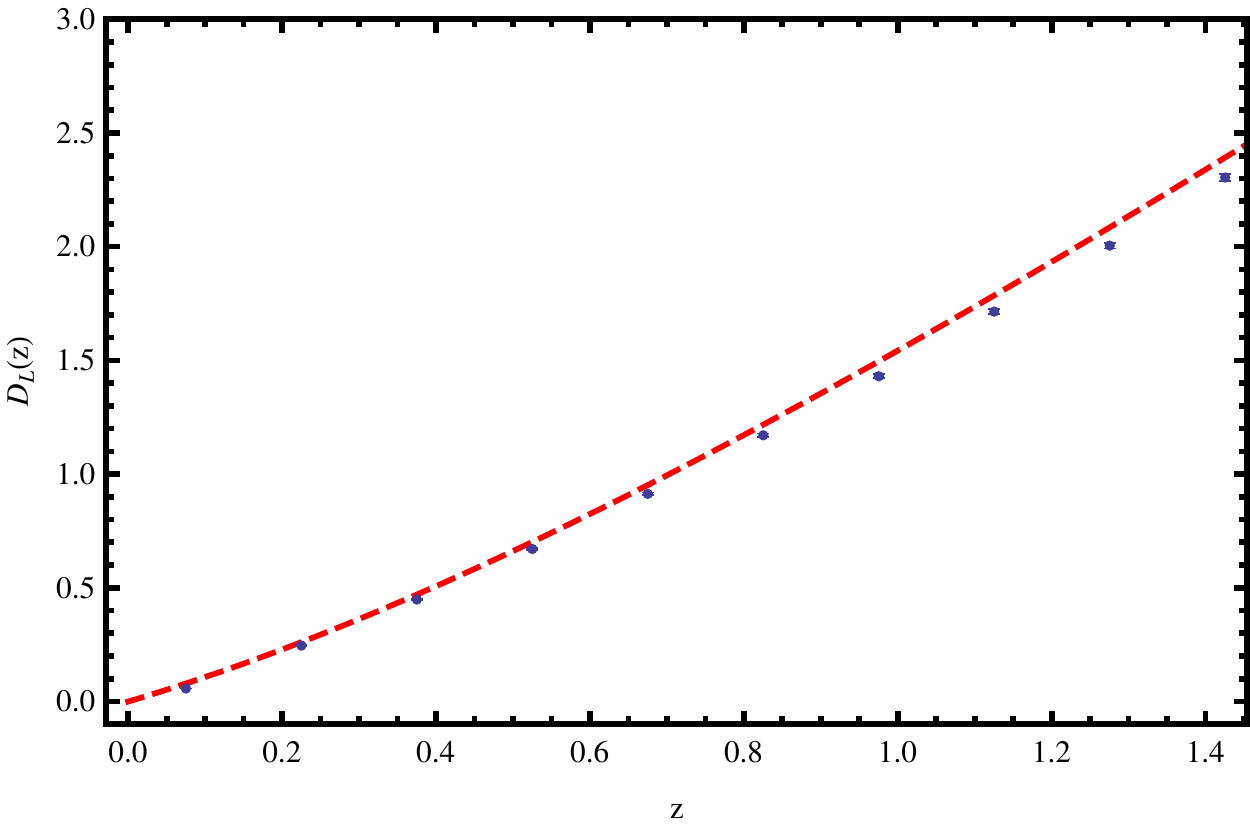}
\includegraphics[scale=0.45]{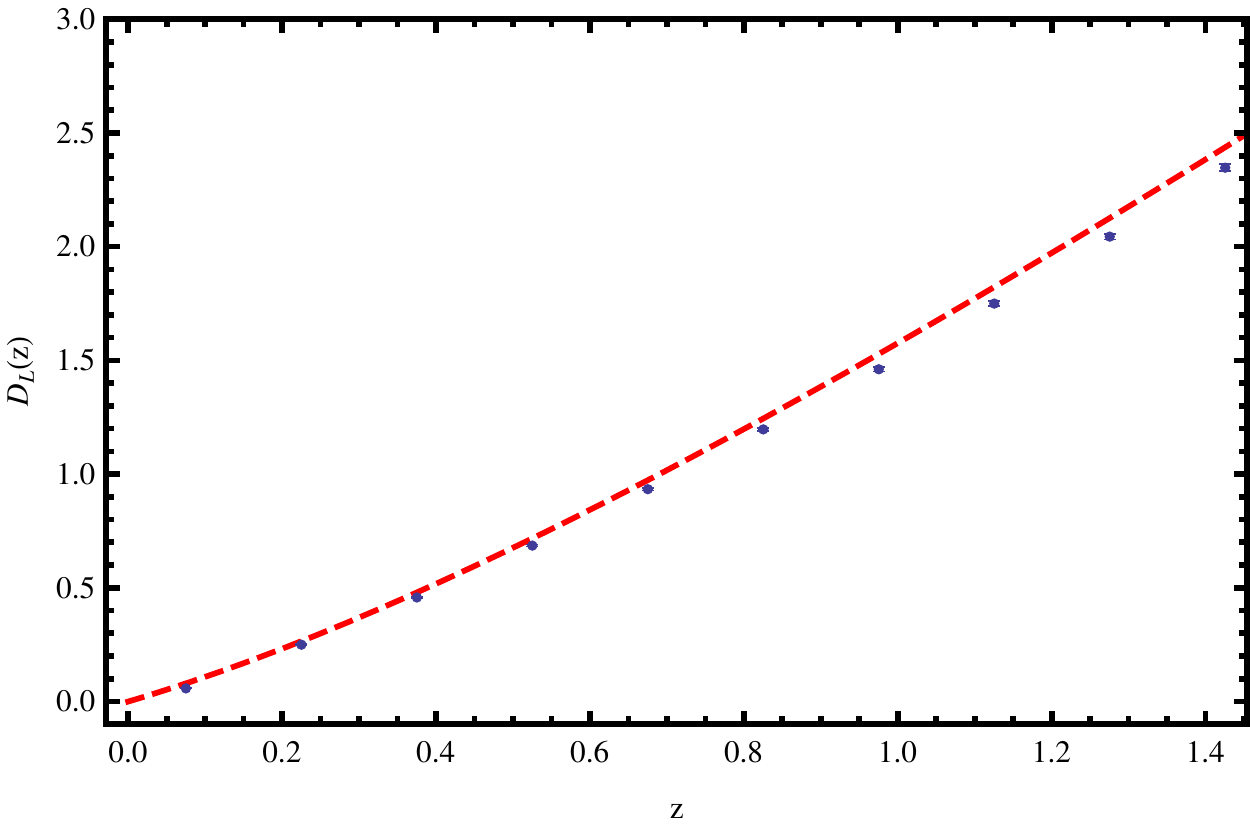}
\includegraphics[scale=0.45]{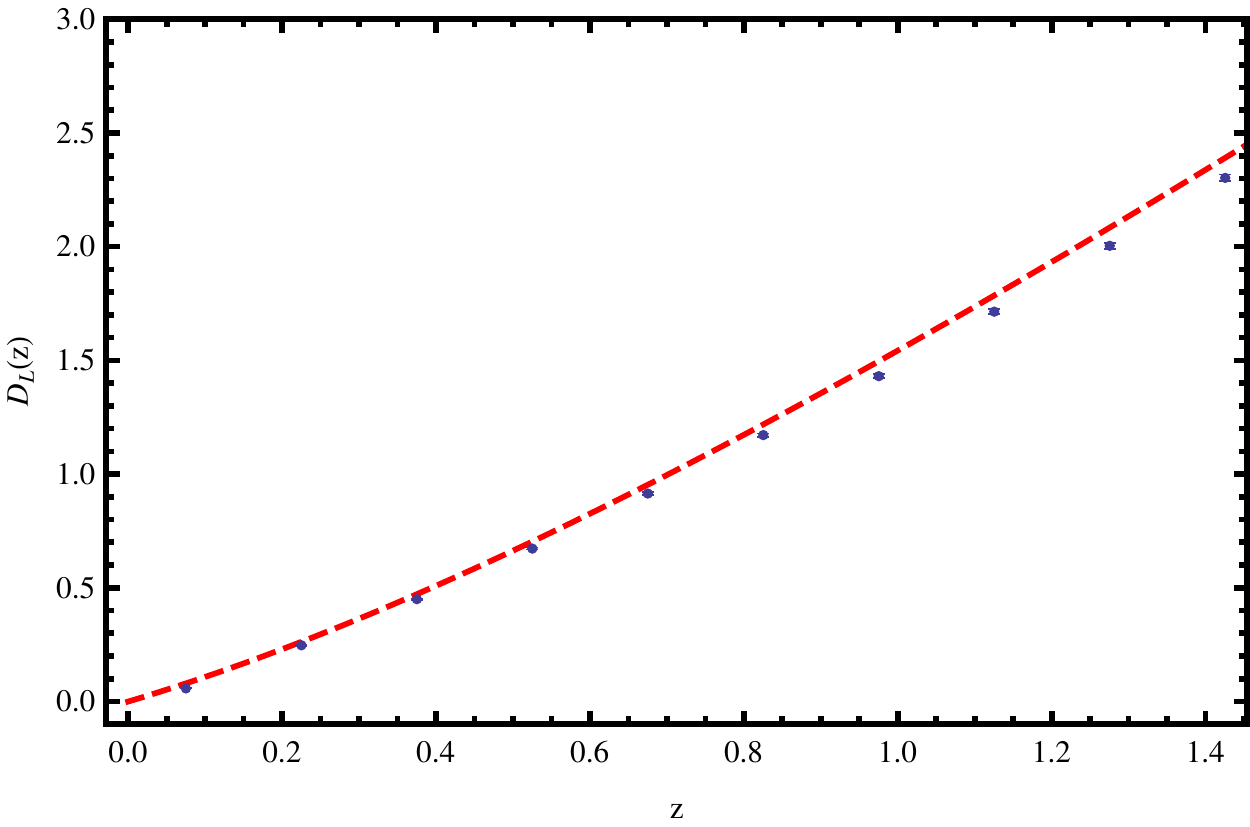}
\caption{Left column refers to the $\Lambda$CDM cosmology, center column to the CPL cosmology and right column to the Hu \& Sawicki $f(R)$ cosmology with $n=1$ and $b=0.1$. In all the plots the red dashed line is the corresponding theoretical functions used to create the mock catalogs. For the $q$ and $w$ parameters (first and second rows) we show the PCA values given by Eqs.~(\ref{eq:pca-values}) and (\ref{eq:pca-errors}), while for $H$ and $D_L$ (third and fourth rows) we show the best-fit values.}
\label{fig:pca}
\end{figure}

In Fig. \ref{fig:flowchart} we present a flowchart that illustrates and clarifies our methodology, while in Fig. \ref{fig:pca} we plot the PCA values of the deceleration parameter $q$ and the equation of state $w$ for the 10 bins (first and second row respectively),
while in Table \ref{tab:pca-q-w} we show their corresponding values and $1\,\sigma$ errors.
In Fig. \ref{fig:pca} we show the best-fit parameters for the Hubble parameter $H$ and $D_{L}$ (third and fourth row), while in Table \ref{tab:bf-H-dL} we show their corresponding values and $1\,\sigma$ errors. In both cases, the different columns correspond to the different cosmologies $\Lambda$CDM, CPL and $f(R)$, as indicated by the labels.

A limitation with some of the methods is that they require values for $\omo$ in order
to get an estimate of $w(z)$, see Eq.~(\ref{eq:w-fromq}) or values of $H_0$ like in the fourth
method where it cannot be estimated by the data or at least marginalized over. In this case,
since we are not supposed to know the true parameters of our cosmology we will use the
Planck best fits $\omo=0.315\pm0.017$ and $H_0=(67.3\pm1.2)\textrm{km} \textrm{s}^{-1} \textrm{Mpc}^{-1}$. In all cases we took care to propagate the errors from $\omo$ and $H_0$ to the derived parameters.

\begin{table}
\begin{centering}\begin{tabular}{|c|c|c|c|c|c|c|}
\hline
& $\Lambda$CDM &  CPL &  $f(R)$ & $\Lambda$CDM &  CPL &  $f(R)$ \tabularnewline
\hline
& \multicolumn{3}{|c|}{\bf{q(z)} } &\multicolumn{3}{|c|}{\bf{w(z)}} \tabularnewline
\hline
$\chi^2_{min}$ & $1844.28$ & $1844.24$ & $1844.35$ & $1846.08$ & $1847.1$ & $1846.88$ \tabularnewline
\hline
$z_{r}$ & $q_n\pm 1\sigma_{q_{n}}$ & $q_n\pm 1\sigma_{q_{n}}$ & $q_n\pm 1\sigma_{q_{n}}$ & $w_n\pm\sigma_{w_{n}}$ & $w_n\pm\sigma_{w_{n}}$ & $w_n\pm\sigma_{w_{n}}$ \tabularnewline
\hline
$0.075$ & $-0.533\pm0.019$ & $-0.690\pm0.020$ & $-0.499\pm0.019$  & $-0.795\pm0.076$ & $-1.197\pm0.083$ & $-0.903\pm0.073$\tabularnewline
\hline
$0.225$ & $-0.478\pm0.056$ & $-0.552\pm0.056$ & $-0.415\pm0.056$ & $-1.117\pm0.111$ & $-1.519\pm0.125$ & $-1.289\pm0.155$  \tabularnewline
\hline
$0.375$ & $-0.323\pm0.114$ & $-0.344\pm0.116$ & $-0.266\pm0.113$ & $-1.156\pm0.166$ & $-1.537\pm0.177$ & $-0.935\pm0.327$\tabularnewline
\hline
$0.525$ & $-0.201\pm0.218$ & $-0.194\pm0.215$ & $-0.131\pm0.215$ & $-1.152\pm0.219$ & $-1.277\pm0.291$ & $-1.181\pm0.565$\tabularnewline
\hline
$0.675$ & $0.019\pm0.386$ & $0.109\pm0.405$ & $0.141\pm0.394$  & $-0.779\pm0.378$ & $-1.160\pm0.479$ & $-0.271\pm0.716$\tabularnewline
\hline
$0.825$ & $0.469\pm0.671$ & $0.517\pm0.696$ & $0.524\pm0.680$ & $0.141\pm0.639$ & $-0.054\pm0.765$ & $0.094\pm1.148$  \tabularnewline
\hline
$0.975$ & $0.520\pm1.111$ & $0.551\pm1.109$ & $0.557\pm1.005$ & $0.857\pm1.128$ & $-0.526\pm0.832$ & $0.553\pm1.591$ \tabularnewline
\hline
$1.125$ & $0.458\pm1.013$ & $0.457\pm1.655$ & $0.428\pm1.285$  & $-0.035\pm2.513$ & $0.801\pm2.155$ & $0.263\pm1.999$ \tabularnewline
\hline
$1.275$ & $0.249\pm3.621$ & $0.406\pm2.661$ & $0.326\pm2.575$  & $0.391\pm4.176$ & $0.665\pm2.798$ & $-0.076\pm4.496$ \tabularnewline
\hline
$1.425$ & $0.321\pm4.784$ & $0.228\pm4.425$ & $0.183\pm4.871$ & $0.023\pm5.716$ & $0.116\pm4.519$ & $0.400\pm8.965$  \tabularnewline
\hline
\end{tabular}\par\end{centering}
\caption{PCA values for $q$ and $w$ and their $1\sigma$ errors for three different cosmologies. Note that while we report the best-fit value of the $\chi^2$ from the MCMC, i.e. the value at the minimum $\chi^2_{min}$, the values of the parameters are the ones that result from the PCA and not the best-fit ones.
\label{tab:pca-q-w}}
\end{table}
%%%%%%%%%%%%%%%%%%

\begin{table}
\begin{centering}\begin{tabular}{|c|c|c|c|c|c|c|}
\hline
& $\Lambda$CDM & CPL &  $f(R)$ & $\Lambda$CDM &  CPL &  $f(R)$ \tabularnewline
\hline
& \multicolumn{3}{|c|}{\bf{H(z)} } &\multicolumn{3}{|c|}{\bf{D$_L$(z)}} \tabularnewline
\hline
$\chi^2_{min}$ & $1866.4$ & $1856.0$ & $1868.1$ & $-$ & $-$ & $-$ \tabularnewline
\hline
$z_{r}$ & $H_n\pm 1\sigma_{H_{n}}$ & $H_n\pm 1\sigma_{H_{n}}$ & $H_n\pm 1\sigma_{H_{n}}$ & $D_L\pm\sigma_{D_L}$ & $D_L\pm\sigma_{D_L}$ & $D_L\pm\sigma_{D_L}$ \tabularnewline
\hline
$0.075$ & $ 1.062\pm0.012$ & $ 1.054\pm 0.012$ & $ 1.055\pm0.012$  & $ 0.057\pm0.0004$ & $ 0.058\pm0.0004$ & $ 0.058\pm 0.0004 $\tabularnewline
\hline
$0.225$ & $ 1.185\pm0.017$ & $ 1.148\pm 0.016$ & $ 1.182\pm0.017$  & $ 0.246\pm0.002$ & $ 0.250\pm0.002$ & $ 0.247\pm 0.002 $\tabularnewline
\hline
$0.375$ & $ 1.274\pm0.026$ & $ 1.242\pm 0.026$ & $ 1.273\pm0.026$  & $ 0.449\pm0.003$ & $ 0.458\pm0.003$ & $ 0.450\pm 0.003 $\tabularnewline
\hline
$0.525$ & $ 1.381\pm0.039$ & $ 1.344\pm0.038 $ & $ 1.384\pm0.040$  & $ 0.671\pm0.004$ & $ 0.686\pm0.004$ & $ 0.672\pm 0.004 $\tabularnewline
\hline
$0.675$ & $ 1.500\pm0.057$ & $ 1.469\pm 0.55$ & $ 1.505\pm0.057$  & $ 0.913\pm0.006$ & $ 0.933\pm0.006$ & $ 0.914\pm 0.006 $\tabularnewline
\hline
$0.825$ & $ 1.697\pm0.084$ & $ 1.667\pm 0.083$ & $ 1.703\pm0.085$  & $ 1.171\pm0.008$ & $ 1.197\pm0.008$ & $ 1.171\pm 0.008 $\tabularnewline
\hline
$0.975$ & $ 1.832\pm0.111$ & $ 1.806\pm 0.110$ & $ 1.839\pm0.112$  & $ 1.430\pm0.009$ & $ 1.461\pm0.009$ & $ 1.430\pm 0.009 $\tabularnewline
\hline
$1.125$ & $ 1.897\pm0.132$ & $ 1.876\pm 0.132$ & $ 1.903\pm0.133$  & $ 1.714\pm0.011$ & $ 1.750\pm0.011$ & $ 1.714\pm 0.011 $\tabularnewline
\hline
$1.275$ & $ 2.241\pm0.205$ & $ 2.225\pm 0.206$ & $ 2.249\pm0.206$  & $ 2.005\pm0.013$ & $ 2.044\pm0.013$ & $ 2.200\pm 0.013 $\tabularnewline
\hline
$1.425$ & $ 1.975\pm0.215$ & $ 1.960\pm 0.216$ & $ 1.980\pm0.216$  & $ 2.304\pm0.015$ & $ 2.347\pm0.015$ & $ 2.302\pm 0.015$\tabularnewline
\hline
\end{tabular}\par\end{centering}
\caption{The best-fit values for $H$ and $d_L$ and their $1\sigma$ errors for three different cosmologies. Note that we do not present the best-fit values for the $\chi^2$ for the luminosity distance $D_L(z)$, as in this case the $\chi^2_{min}$ is too large, since as can be easily seen by inspecting the definition of the chi-square, the value of $D_L(z)$ in each bin will be constant, thus over(under)estimating the difference between the theoretical value and the measured one and as a result affecting the $\chi^2$. The effect of the discontinuity is also shown in Fig. \ref{fig:dlvsdl}, where the case of the piecewise-constant $D_L$ is compared to $D_L$ for the piecewise-constant $q$. \label{tab:bf-H-dL}}
\end{table}
%%%%%%%%%%%%%%%%%%

Schematically, the difficulties of the different methods are
\begin{itemize}
 \item Piecewise-constant $q_n$: when we want to extract information on the EOS parameter $w$, we need to assume a value for the matter density $\omo$, for which we use the Planck prior mentioned above; this is because the expression of the luminosity distant Eq.~(\ref{eq:dl-q}) does not depend on $\omo$, consequently we cannot use the chains to estimate the best-fit value of the matter density. On the other hand, the Hubble parameter $H_n$ and the luminosity distance $d_{L,n}$ do not depend on $\omo$.
\item Piecewise-constant $w_n$: $\omo$ is a free parameter and we can find the best-fit value of $\omo$ directly from the chain along with the values of $w_n$. This value for $\omo$ will be propagated to all the other parameters, $q_n$, $H_n$ and $d_{L,n}$, as they all depend on the matter density $\omo$.
\item Piecewise-constant $H_n$: when we want to extract information on the EOS parameter $w$, we need to assume a value for the matter density $\omo$, for which we use the Planck prior mentioned above. On the other hand, the deceleration parameter $q_n$ and the luminosity distance $d_{L,n}$ do not depend on $\omo$.
However, using this method to find the Hubble parameter, we need to assume a value for the Hubble constant $H_0$.
\item Piecewise-constant $\mu_n$: in this case the deceleration parameter and the Hubble parameter do not depend on $\omo$, but the EOS parameter does so the matter density parameter needs to be propagated. Also using this method, in order to find the best fit of the quantities $d_{L,n}$ we need to use a prior for the Hubble constant $H_0$.
\end{itemize}

As can be seen in Tables.~\ref{tab:pca-q-w} and \ref{tab:bf-H-dL} (and also from Fig.~\ref{fig:pca}), the method that gives the least errors is the direct measurement of the luminosity distance $D_{L}$ (which is the dimensionless luminosity distant),
where the errors are of the order of $0.5\%$ in average with respect to their corresponding measurements. We should remind that we have not marginalized over the Hubble constant $H_0$, for which we assumed it as constant. This result is quite obvious since SnIa directly measure the distance modulus $\mu$ which is connected to the luminosity distance; this is also reflected in Table ~\ref{tab:risk-LCDM} where the best reconstruction of the derived parameters comes from the direct measurement of the luminosity distance.

Finally, we should stress that the values shown in Table \ref{tab:bf-H-dL} correspond to the dimensionless luminosity distance $D_L$, so the errors do not have any units. Regarding the small value of the errors, these can be explained by understanding how the propagation of the errors occurs. The error on $D_L$ will be given by the derivative of $D_L$ with respect to $\mu_i$, see (\ref{eq:DLi}), times the error of the best-fit distance modulus, i.e. $\sigma_{D_L}^2=\left(\partial_\mu D_L \sigma_{\mu}\right)^2$, but $\sigma_{\mu}$ is given by (\ref{muerrors}) and (\ref{eq:SnApp}) $\sigma_\mu^2\sim\sigma_i^2/200$, where $\sigma_i=0.1$ for all points, so that $\sigma_\mu\sim0.007$ and $\sigma_{D_L} \sim10^{-3}$.

\subsection{Comparison to the different parameters}
In this section we will compare the different methods and we will also study
their advantages and weakness with respect to each other.

Let us define two quantities, called the bias and variance, \cite{huterstark} and \cite{wasserman}:
\ba
\textrm{bias}&=&\sum_{i=1}^{n_{tot}}(y(z_i)-y_{real}(z_i))^2\\
\textrm{variance}&=&\sum_{i=1}^{n_{tot}}\sigma(y(z_i))^2,
\ea
where $y(z_i)$ are the reconstructed parameters in each bin i.e. $w,q,H,d_L$, $y_{real}(z_i)$
are the ``real" values of the parameters, and $n_{tot}$ is the number of points.
The bias tells us how different are the reconstructed parameters from the real ones,
while the variance tells us how big are the errors.

Then we also define the risk as the sum of the two
\be
\textrm{risk}=\textrm{bias}+\textrm{variance}, \label{riskstat}
\ee
so that reconstruction methods that give results closer to the real cosmology and have smaller errors, will have a smaller value for the risk. We prefer to use the risk rather than the usual $\chi^2$ analysis for two reasons: first, the risk is precisely aimed at measuring the closeness of an estimated quantity to the corresponding theoretical function taking in consideration also the errors of the estimated quantity; second, as reported in details in the Appendix B, the $\chi^2$ analysis fails to describes the goodness of the data in this analysis. In this case, the risk better describes the goodness of the reconstructed quantities as we are not interested to find the best fit to the data but rather to find the reconstructed quantities that give the original cosmology. Let us make an example to clarify the problem with the $\chi^2$. Imagine we have two different cases: in the first case, the reconstructed quantity is far from the real one but it has small errors, and in the second case the reconstructed quantity is very close to the real one but it has large errors. The two reconstructions might give the same risk implying that they are equally bad. If instead we use the $\chi^2$, then the second case would be preferred with respect to the first one, thus giving a biased result.  Since our aim is to find the best reconstruction of the curve with the smallest errors, the second case is not to be preferred to the first one.

For the sake of completeness, we also evaluated the $\chi^2$ for all the four reconstructions for the three different cosmologies in Appendix B. However, in this context, these values should be not taken seriously because the $\chi^2$ fails to describe the closeness of reconstructed quantities, as previously reported. The results for the risk for the three cosmologies are shown in Tables \ref{tab:risk-LCDM}, \ref{tab:risk-CPL} and \ref{tab:risk-f_R} respectively. The columns indicate the methods use to fit the data, while the rows indicate the reconstructed parameters. For example, if we want the value of the risk parameter for the luminosity distance $D_L$ and for the $q(z)$ piecewise-constant method, we have to pick the element in the second column and fourth row, e.g. in Table \ref{tab:risk-LCDM} that value would be 0.056.

If we choose one parameter of interest (say $w$) and ask what is the best method to reconstruct this parameter, then we find that the best method is always the luminosity distance piecewise-constant scheme, followed by $w$. Between the two PCA methods considered in this analysis ($q$ and $w$), $w$ is better at reconstructing both $q$ and $w$.

The problem here is that if we want to reconstruct $w$ from $q$, then the error propagation formula is
\be
\sigma_w = \sqrt{\left(\frac{\partial w}{\partial q}\right)^2\sigma_q^2+
\left(\frac{\partial w}{\partial\omo}\right)^2\sigma_{\omo}^2}\,.
\label{eq:errwfromq}
\ee
Both terms in Eq.~(\ref{eq:errwfromq}) are proportional to $(1+z)^{-3 w}$ which is an increasing
function with redshift; however, the small value of $\sigma_{\omo}$ washes away the information of
the matter density and the only contribution comes from the $q_n$, leading to $\sigma_w\gg\sigma_q$.
This also explains the values of the risk of about $215$ (first row and second column)
in Table~\ref{tab:risk-LCDM}.

The same discussion can be applied to the case when we have $H$ and we want to reconstruct $w$. In this case we have
\be
\sigma_w = \sqrt{\left(\frac{\partial w}{\partial H}\right)^2\sigma_H^2+
\left(\frac{\partial w}{\partial\omo}\right)^2\sigma_{\omo}^2}\,.
\label{eq:errwfromh}
\ee
The derivative in terms of $H$ scales as $\partial w/\partial H \sim (1+z)^{3/2}/\ln(1+z)$, whereas the one in terms of $\omo$ scales as $\partial w/\partial \omo \sim (1+z)^{-3w}/\ln(1+z)$.
Clearly, they both contribute equally to the final errors and they are much larger than unity, with the first term being dominant at low redshifts, whereas the second term dominates at high redshifts. Also, we should note that the equation used for deriving $w$ from $H$ is similar to the equation used to derive $w$ from $q$, something which explains the magnitude of the errors (see for instance first row and third column in Table~\ref{tab:risk-LCDM}.

Finally, it should be stressed that all methods have certain limitations. For example, as mentioned in the previous paragraph, a value for $\omo$ is required in order to get an estimate of $w(z)$, see Eq.~(\ref{eq:w-fromq}) or values of $H_0$, like in the fourth method, are needed since $H_0$ cannot be estimated by the data or at least marginalized over. As we are interested in making a ``blind" comparison of the methods, we are not supposed to know the true values of the parameters of our underlying real cosmology, so we used the Planck best fits previously defined.
In all cases we took care to propagate the errors from $\omo$ and $H_0$ to the derived parameters.

As a final remark, it is interesting to notice that the errors for the four reconstructed quantities $q(z)$, $w(z)$, $H(z)$, $d_L(z)$ increase with redshift. The reasons are the following: for the deceleration parameter $q(z)$ and the EOS parameter $w(z)$ we performed the PC analysis and the transformation matrix $\tilde{W}_{ij}$ is composed by the eigenvectors of the Fisher matrix whose rows, that are the eigenvectors, have been ordered according to the corresponding eigenvalues. In practice, the first row is the eigenvector with the smallest eigenvalue and the last row is the eigenvector with the largest eigenvalue, see Sec. \ref{sec:background}.

For the Hubble parameter $H(z)$ and the luminosity distant $d_L(z)$ the errors increase with redshift simply because the propagation formula is a linear function of the redshift.

%%%%%%%%%%%%%%%%
\begin{table}[t!]
\begin{centering}\begin{tabular}{|c|c|c|c|c|}
\hline
\multicolumn{5}{|c|}{The risk for $\Lambda$CDM}  \tabularnewline
\hline
 \backslashbox{{\small Derived param.}}{Piecewise} & $w$ & $q$ & $H$ & $D_L$  \tabularnewline
\hline
$w$ & $62.514$ & $215.307$ & $267.176$ & $18.784$  \tabularnewline
\hline
$q$ & $16.373$ & $42.341$ & $30.965$ & $2.903$  \tabularnewline
\hline
$H$  & $41.751$ & $0.597$ & $0.265$ & $0.038$ \tabularnewline
\hline
$D_L$  & $0.205$ & $0.056$ & $0.032$ & $0.030$ \tabularnewline
\hline
\end{tabular}\par\end{centering}
\caption{The values of the risk parameter for the $\Lambda$CDM cosmology. The rows indicate the derived parameters, while the columns indicate the piecewise method.
\label{tab:risk-LCDM}}
%\end{table}
%%%%%%%%%%%%%%%%%%%
%
%%%%%%%%%%%%%%%%%
%\begin{table}
\begin{centering}\begin{tabular}{|c|c|c|c|c|}
\hline
\multicolumn{5}{|c|}{The risk for CPL}  \tabularnewline
\hline
 \backslashbox{{\small Derived param.}}{Piecewise} & $w$ & $q$ & $H$ & $D_L$  \tabularnewline
\hline
$w$ & $39.972$ & $278.300$ & $302.870$ & $23.510$  \tabularnewline
\hline
$q$ & $13.730$ & $32.855$ & $31.572$ & $3.099$  \tabularnewline
\hline
$H$  & $37.665$ & $0.532$ & $0.264$ & $0.036$ \tabularnewline
\hline
$D_L$  & $0.134$ & $0.063$ & $0.034$ & $0.031$ \tabularnewline
\hline
\end{tabular}\par\end{centering}
\caption{The values of the risk parameter for the CPL cosmology. The rows indicate the derived parameters, while the columns indicate the piecewise method.
\label{tab:risk-CPL}}
%\end{table}
%%%%%%%%%%%%%%%%%%%
%%%%%%%%%%%%%%%%%
%\begin{table}
\begin{centering}\begin{tabular}{|c|c|c|c|c|}
\hline
\multicolumn{5}{|c|}{The risk for $f(R)$}  \tabularnewline
\hline
 \backslashbox{{\small Derived param.}}{Piecewise} & $w$ & $q$ & $H$ & $D_L$  \tabularnewline
\hline
$w$ & $112.263$ & $98.073$ & $260.484$ & $18.038$  \tabularnewline
\hline
$q$ & $24.507$ & $34.166$ & $31.149$ & $3.564$  \tabularnewline
\hline
$H$  & $88.713$ & $0.615$ & $0.268$ & $0.038$ \tabularnewline
\hline
$D_L$  & $0.300$ & $0.046$ & $0.032$ & $0.030$ \tabularnewline
\hline
\end{tabular}\par\end{centering}
\caption{The values of the risk parameter for the $f(R)$ cosmology. The rows indicate the derived parameters, while the columns indicate the piecewise method.
\label{tab:risk-f_R}}
\end{table}

\section{Conclusions}\label{sec:conclusions}
In this paper we compared the four different methods that can be used to analyze the type Ia supernovae (SnIa) data, i.e. use different piecewise-constant functions, such as: the dark energy equation of state $w(z)$, the deceleration parameter $q(z)$, the Hubble parameter $H(z)$ and finally the luminosity distance $d_L(z)$. These four quantities cover all main aspects of the accelerating Universe, i.e. the phenomenological properties of dark energy, the expansion rate (first and second derivatives) of the Universe and the observations themselves.

For the first two cases we also performed principal component analysis (PCA) so as to decorrelate the parameters, while for the last two cases we used a set of novel analytic expressions for the best fit. We derived the equations for the PCA for the two methods ($w$ and $q$), while for the other two ($H$ and $D_L$) we used their best fits as for the former we found that the parameters are very highly correlated so that a linear transformation, i.e. the PCA, cannot decorrelate them and that due to the degeneracy the MCMC also fails, while for the latter we found that the covariance matrix is already diagonal.

In order to test the methods we created sets of mock SnIa data (2000 points uniformly distributed in redshift $z\in[0,1.5]$) for three fiducial cosmologies, the cosmological constant model ($\Lambda$CDM), a linear expansion of the dark energy equation of state parameter $w(a)=w_0+w_a(1-a)$ and the Hu-Sawicki $f(R)$ model. Then we fitted the piecewise schemes on the mock data, either with a MCMC or by using the analytic formulas, and we found the best-fit parameters, to which we applied the PCA for the $w$ and $q$ methods.

In the last step of our methodology, we compared the four different forms of the PCA using the risk statistic defined in Eq.~(\ref{riskstat}) and ranked the methods accordingly. The final results of our analysis can be seen in Tables \ref{tab:risk-LCDM}, \ref{tab:risk-CPL} and \ref{tab:risk-f_R}. These Tables can help us answer the question: Given a parameter of interest, what is the best piecewise-constant method to reconstruct it?

To answer this question it is best to initially focus on the two mainstream approaches for the PCA, i.e. the $w(z)$ and $q(z)$ piecewise-constant methods. By inspecting Tables \ref{tab:risk-LCDM}, \ref{tab:risk-CPL} and \ref{tab:risk-f_R} we see that, given a parameter of interest, i.e. moving horizontally on the Tables, then the best piecewise-constant scheme is always $w(z)$.

If we also take the other two parameters into account, then we see that the best piecewise-constant method overall is $d_L(z)$, i.e. traditional binning, due to the fact that in this case the errors on the best-fit parameters are significantly smaller than in the other cases. In general, all methods suffer from a few limitations, for example using the $q(z)$ scheme implies that in order to get a constraint on $w(z)$ we have to assume a value for $\omo$, as that cannot be estimated from the data alone.

Overall, the novelty of our analysis is twofold: we performed a systematic comparison of all the methods and subsequently tested them against mock data and against each other, for the first time in the literature and second, we presented several new analytical expressions for all four different forms of piecewise-constant methods. Our analysis will be immensely useful with the upcoming surveys that will collect a plethora of new data that will have to be analyzed in a systematic fashion and their cosmological information extracted. Moreover, it is rather straightforward to include other kinds of data as well and get even more stringent constraints on the parameters, but also to use more realistic mocks to see which of the methods performs the best in more realistic scenarios.

%%%%%%%%%%%%%%%%

\section*{Acknowledgments}
We would like to thank K.~Rolbiecki and the anonymous referee for useful suggestions to the manuscript. The authors acknowledge financial support from the Madrid Regional Government (CAM) under the program HEPHACOS S2009/ESP-1473-02, from MICINN under Grant No. AYA2009-13936-C06-06 and Consolider-Ingenio 2010 PAU (CSD2007-00060), as well as from the European Union Marie Curie Initial Training Network UNILHC Grant No. PITN-GA-2009-237920. We also acknowledge the support of the Spanish MINECO's Centro de Excelencia Severo Ochoa programme under Grant No. SEV-2012-0249.

\appendix
\section{Equations for the PCA}
\subsection{The deceleration parameter \label{app-PCAq}}
We start piecewise-constant the deceleration parameter as
\be
q(z) = \sum_{i=1}^{n}q_i\theta(z_i),
\ee
where $q_i$ are constant in each redshift bin $z_i$ and $\theta(z_i)$ is the theta function, i.e.
$\theta(z_i) = 1$ for $z_{i-1}< z \leq z_{i}$ and $0$ elsewhere.
The general expression for the deceleration parameter is
\be
1+q(z)=\frac{d \ln(H(z))}{d\ln(1+z)}\,.
\ee
The last equation can be inverted to find the Hubble parameter:
\be
\ln(H(z)/H_0)=\int_0^z\frac{1+q(x)}{1+x}dx,
\ee
or
\be
H(z)/H_0 = e^{I(z)}\,\hspace{1cm} {\rm where}\hspace{1cm} I(z) =\int_0^z\frac{1+q(x)}{1+x}dx\,.
\ee
For $z \in(z_{i-1},z_i]$ and using the fact that $q$ is constant in each bin, we can break the integral $I(z)$ in parts as
\bea
I(z)&=&\int_0^{z_1}(...)+\int_{z_1}^{z_2}(...)+...+\int_{z_{i-1}}^{z}(...)\nn\\
&=&(1+q_1)\ln(1+x)|_0^{z_1}+(1+q_2)\ln(1+x)|_{z_1}^{z_2}+...+(1+q_i)\ln(1+x)|_{z_{i-1}}^{z}\nn\\
&=&(1+q_1)\ln(1+z_1)+(1+q_2)\ln\left(\frac{1+z_2}{1+z_1}\right)+...+(1+q_i)\ln\left(\frac{1+z}{1+z_{i-1}}\right).
\eea
Grouping the constant terms, the Hubble parameter can then be written, for $z$ in the nth bin, as
\be
H_n(z) = H_0 b_n \left(1+z\right)^{1+q_n}
\ee
where the coefficient $b_n$ is
\be
b_n = \prod_{j=1}^{n-1}\left(1+z_j\right)^{q_j-q_{j+1}}\,.
\ee
We can now follow a similar procedure to calculate the luminosity distance.
Using the definition of the luminosity distance along with the previous equations we have
\bea
d_L(z)&=&\frac{c}{H_0} (1+z) \int_0^z\frac{1}{H(x)/H_0} \rmd x \nn \\
&=&\frac{c}{H_0} (1+z)\left( \int_0^{z_1}(...)+\int_{z_1}^{z_2}(...)+...+\int_{z_{i-1}}^{z}(...)\right)\nn\\
&=&\frac{c}{H_0} (1+z) \left( \frac{1-(1+z_1)^{-q1}}{b_1 q_1}+\frac{(1+z_1)^{-q_2}-(1+z_2)^{-q_2}}{b_2 q_2}+...+ \frac{(1+z_{i-1})^{-q_i}-(1+z)^{-q_i}}{b_{i} q_i}\right).
\eea
Collecting the constant terms, the latter can be written as
\be
d_{L,n}(z)=\frac{c}{H_0}\left(1+z\right)\left[f_n-\frac{\left(1+z\right)^{-q_n}}{b_n q_n}\right],
\ee
where
\be
f_n =  \frac{\left(1+z_{n-1}\right)^{-q_n}}{b_n q_n}+\sum_{j=1}^{n-1}\frac{\left(1+z_{j-1}\right)^{-q_j}-
\left(1+z_{j}\right)^{-q_j}}{b_j q_j}\,,
\ee
being $z_0=0$.

\subsection{The dark energy equation of state parameter \label{app-PCAw}}

We now want to apply the PCA directly to the dark energy EOS parameter $w(z)$. As previously done for $q$, we rewrite $w(z)$ as
\be
w(z) = \sum_{i=1}^{n_{tot}}w_i\theta(z_i),
\ee
where $w_i$ are constant in each redshift bin $z_i$, $n_{tot}$ is the total number of bins and
$\theta(z_i)$ is the theta function, i.e. $\theta(z_i) = 1$ for $z_{i-1}< z \leq z_{i}$ and $0$ elsewhere.Using the energy-momentum conservation $\nabla_\mu T^{\mu\nu}=0$ for an ideal fluid with equation of state $w$ we get the equation
\be
\dot{\rho}_{DE}+3(1+w) H\rho_{DE}=0,
\ee
which can be solved and written in terms of the redshift $z$. Then, the general expression for the DE density in terms of a time-dependent $w(z)$ is
\be
\rho_{DE}(z)=\rho_{DE}(z=0)e^{3\int_0^z\frac{1+w(z')}{1+z'} dz'},
\ee
and this can be rewritten as
\be
\frac{1}{3}\ln\left(\frac{\rho_{DE}(z)}{\rho_{DE}(z=0)}\right)=I(z), \label{eq:rhoDE1}
\ee
where
\be
I(z)=\int_0^z\frac{1+w(x)}{1+x}dx \label{eq:rhoDE2}
\ee
Proceeding like before, for $z \in(z_{i-1},z_i]$ and using the fact that $w$ is constant in each bin,
we can break the integral $I(z)$ in parts as
\bea
I(z)&=&\int_0^{z_1}(...)+\int_{z_1}^{z_2}(...)+...+\int_{z_{i-1}}^{z}(...)\nn\\
&=&(1+w_1)\ln(1+x)|_0^{z_1}+(1+w_2)\ln(1+x)|_{z_1}^{z_2}+...+(1+w_i)\ln(1+x)|_{z_{i-1}}^{z}\nn\\
&=&(1+w_1)\ln(1+z_1)+(1+w_2)\ln\left(\frac{1+z_2}{1+z_1}\right)+...+(1+w_i)\ln\left(\frac{1+z}{1+z_{i-1}}\right).
\eea
Grouping the constant terms, the dark energy density can then be written, for $z$ in the nth bin, as
\be
\rho_{DE}(z,n)=\rho_{DE}(z=0) c_n \left(1+z\right)^{3(1+w_n)},
\ee
where the coefficient $c_n$ is
\be
c_n = \prod_{j=1}^{n-1}\left(1+z_j\right)^{w_j-w_{j+1}}\,,
\ee
and obviously for $n=1$ we have $c_1=1$, since by definition $\prod_{j=1}^{0}(...)\equiv 1$.
Then, if we also include matter, the Hubble parameter can be written as
\be
H(z,n)^2/H_0^2=\omo (1+z)^3+(1-\omo)c_n \left(1+z\right)^{3(1+w_n)}.
\ee
Since $c_1=1$ we have that for $z=0$, i.e. for the first bin or $n=1$, $H(z=0,n=1)=H_0$ as expected. We can now follow a similar procedure to calculate the luminosity distance.
Using the definition of the luminosity distance along with the previous equations we have
\bea
d_L(z)&=&\frac{c}{H_0} (1+z) \int_0^z\frac{1}{H(x)/H_0} \rmd x \nn \\
&=&\frac{c}{H_0} (1+z)\left( \int_0^{z_1}(...)+\int_{z_1}^{z_2}(...)+...+\int_{z_{i-1}}^{z}(...)\right).
\eea
However, in this case the integrals are significantly more complicated due to the presence of the matter term. So, for the $i\textrm{th}$ term we have:
\ba
&&d_i(z_i,z_{i-1})\equiv\int_{z_{i-1}}^{z_i}\frac{\rmd z}{\sqrt{\omo (1+z)^3+(1-\omo)c_i \left(1+z\right)^{3(1+w_i)}}}=\nn\\ &&
-\frac{2}{\omo^{1/2}} \left\{ \frac{ _2F_1\left[\frac{1}{2}, -\frac{1}{6w_i}, 1-\frac{1}{6w_i};
-c_i\frac{1-\omo}{\omo}\left(1+z_i\right)^{3w_i}\right]}{\sqrt{1+z_i}}
 - \frac{ _2F_1\left[\frac{1}{2}, -\frac{1}{6w_i}, 1-\frac{1}{6w_i};
-c_i\frac{1-\omo}{\omo}\left(1+z_{i-1}\right)^{3w_i}\right]}{\sqrt{1+z_{i-1}}} \right\}, ~~
\ea
where ${}_2F_1(a,b;c;z)$ is a hypergeometric function defined by the series
\be
{}_2F_1(a,b;c;z)\equiv \frac{\Gamma(c)}{\Gamma(a)\Gamma(b)}\sum^{\infty}_{n=0}\frac{\Gamma(a+n)\Gamma(b+n)}{\Gamma(c+n)n!}z^n
\ee
on the disk $|z|<1$ and by analytic continuation elsewhere; see Ref.~ \cite{handbook} for more details. Now, if we sum up all the terms, the luminosity distance becomes:
\be
d_{L,n}(z)=\frac{c}{H_0} (1+z)\left(d_n(z,z_{n-1})+\sum_{i=1}^{n-1}d_i(z_i,z_{i-1})\right),
\ee
where the last term in the parentheses is just a constant and as always we assume $z_0=0$.
Finally, we have also checked numerically that the expressions above give the correct results.

\subsection{The Hubble parameter $H(z)$ \label{app-PCAh}}

Here we write explicitly the derivation of the cosmological parameters starting from the binned Hubble parameter  $H(z)$. Let us write the Hubble parameter as
\be
H(z)/H_0 = \sum_{i=1}^{n_{tot}}h_i\theta(z_i),
\ee
where $h_i$ are constant in each redshift bin $z_i$, $n_{tot}$ is the total number of bins and
$\theta(z_i)$ is the theta function, i.e. $\theta(z_i) = 1$ for $z_{i-1}< z \leq z_{i}$ and $0$ elsewhere. Using the definition of the luminosity distance along with the previous equations we have
\ba
d_L(z,n)&=& \frac{c}{H_0} (1+z) \left(\int_0^z\frac{1}{H(x)/H_0} \rmd x \right) \nn \\
&=&\frac{c}{H_0} (1+z) \left( \int_0^{z_1}(h_1^{-1} \rmd z)+\int_{z_1}^{z_2}(h_2^{-1} \rmd z)+...+\int_{z_{n-1}}^{z}(h_n^{-1} \rmd z)\right)\nn\\
&=&\frac{c}{H_0} (1+z) \left( h_1^{-1} z_1+h_2^{-1} (z_2-z_1)+...+h_n^{-1}(z-z_{n-1})\right)\nn\\
&=&\frac{c}{H_0} (1+z) \left(\sum_{i=1}^{n-1}z_i(h_i^{-1}-h_{i+1}^{-1})+h_n^{-1} z\right)\nn\\
&=&\frac{c}{H_0} (1+z) \left(g_n+h_n^{-1} z\right),
\ea
where we have defined the constants $g_n\equiv\sum_{i=1}^{n-1}z_i(h_i^{-1}-h_{i+1}^{-1})$.

In what follows we will focus on the case of bins with constant size,
i.e. $z_i-z_{i-1}=dz$ so that $g_n\equiv\sum_{i=1}^{n}\hti_i - n \hti_n$,
but our results can easily be generalized for bins of different sizes as well as well.

Now, we transform the data from the distance modulus $\mu_i$ to
$f_i=\frac{1}{1+z_i} 10^{\frac{\mu_i-25}{5}}$.
Then, the theoretical value is $f_{th}(z,n)=\alpha\left(c_n+\tilde{h}_n z\right)$,
where $\tilde{h}_n=h_n^{-1}$ and $\alpha=\frac{2997.9}{h}$ and the chi square can be written as
\be
\chi^2=\sum_{i=1}^N\left(\frac{f_i-f_{th}(z_i,n)}{\sigma_{i}}\right)^2,\label{chi2f}
\ee
where the errors were found by standard error propagation
$\sigma_{i}^2=\left(\frac{\partial f_i}{\partial \mu_i}\right)^2\sigma_{\mu,i}^2$.
The advantage of this method is that the chi square of Eq.(\ref{chi2f}) is quadratic with respect to the parameters $\tilde{h}_n$ and can be minimized analytically. At this point it is convenient to define the following quantities:
\ba
S&=& \sum_{i=1}^N\frac{1}{\sigma_i^2}, ~~~S_{z^2} = \sum_{i=1}^N\frac{z_i^2}{\sigma_i^2},\\
S_{f^2} &=& \sum_{i=1}^N\frac{f_i^2}{\sigma_i^2},~~~S_{fz} = \sum_{i=1}^N\frac{f_i z_i}{\sigma_i^2},
\ea
and
\ba
S_n&=& \sum_{j=1,\textrm{n bin}}\frac{1}{\sigma_j^2},~~~S_{f_n} = \sum_{j=1,\textrm{n bin}}\frac{f_j}{\sigma_j^2},~~~S_{fz_n} = \sum_{j=1,\textrm{n bin}}\frac{f_j z_j}{\sigma_j^2},\\S_{z_n} &=& \sum_{j=1,\textrm{n bin}}\frac{z_j}{\sigma_j^2},~~~
S_{z^2_n} = \sum_{j=1,\textrm{n bin}}\frac{z^2_j}{\sigma_j^2},
\ea
where $\sum_{j=1,\textrm{n bin}}$ is meant to sum over only those points
in the $n^{\textrm{th}}$ bin, by which it follows that
\be
\sum_{n=1}^{n_{tot}}S_n=S\,,
\ee
where $n_{tot}$ is the number of bins.

With these definitions it is easy to minimize the $\chi^2$ analytically,
following the methodology of Refs.~\cite{press92} and \cite{Nesseris:2012cq}.
The first step is to expand the $\chi^2$ as follows:
\ba
\chi^2&=& \sum_{i=1}^N\frac{1}{\sigma_i^2}\left(f_i^2+\alpha^2 c_n^2+\alpha^2\hti_n^2 z_i^2-2 \alpha f_i c_n-2f_i \alpha \hti_n z_i+2c_n \hti_n \alpha^2 z_i\right)\nn\\
&=&S_{f^2}+\sum_{n=1}^{n_{tot}}\left(\alpha^2 c_n^2S_n+\alpha^2 \hti_n^2 S_{z^2_n}-2\alpha c_n S_{f_n}-2\alpha\hti_n S_{fz_n}+2\alpha^2 c_n \hti_n S_{z_n}\right).
\ea
Now we can define the matrix
\be
A_{nm}\equiv\frac{\partial c_n}{\partial \hti_m}=dz\left(\sum_{k=1}^{n}\delta_{km} -n \delta_{nm}\right).
\ee
Then, the first derivatives of the $\chi^2$ are:
\ba
\beta_k\equiv\frac12\partial_k \chi^2=\hti_k \alpha^2 S_{z^2_k}-\alpha S_{fz_k}+c_k \alpha^2 S_{z_k}+\sum_{n=1}^{n_{tot}}A_{nk}\left(c_n \alpha^2 S_n-\alpha S_{f_n}+\hti_n \alpha^2 S_{z_n}\right),
\ea
while the second derivatives, i.e. the Fisher matrix evaluated at the best fit, are
\ba
\tilde{F}_{kl}&\equiv& \frac12\partial^2_{kl}\chi^2|_{min}=\partial_l \beta_k \nn\\
&=&\alpha^2\left(\delta_{kl} S_{z^2_k}+A_{kl}S_{z_k}+A_{lk}S_{z_l}+\sum_{n=1}^{n_{tot}}A_{nk} A_{nl} S_n\right).\label{fisherH}
\ea
If we define the matrices:
\ba
B_{kl}&=&\alpha^2\left(\delta_{kl} S_{z^2_k}+A_{kl}S_{z_k}+A_{lk}S_{z_l}\right)\nn \\
D_{nk}&=&A_{nk} S_n^{1/2},
\ea
then the Fisher matrix can be written as
\ba
\tilde{F}=B+D^T\,D,
\ea
while the covariance matrix is
\ba
\tilde{C}=\tilde{F}^{-1}=B^{-1}-B^{-1}D^T\left(I+D B^{-1} D^T\right)^{-1}D B^{-1},
\ea
where the last equation comes from considering the inverse of a sum of matrices, see Ref.~\cite{matinv} for details.

As mentioned earlier, in this case the $\chi^2$ is quadratic with respect to $\hti$, so we can use the methodology of Refs.~\cite{press92} and \cite{Nesseris:2012cq}. Clearly, in this case we can write the $\chi^2$ as
\be
\chi^2=\chi^2_{min}+(\hti-\hti_{min})_i \tilde{F}_{ij}(\hti-\hti_{min})_j,
\ee
which means that
\ba
\beta_k=\tilde{F}_{kj}(\hti-\hti_{min})_j
\ea
and that the best-fit parameters and the minimum $\chi^2$are
\ba
\hti_{min,j}&=&-\tilde{F}_{jk}^{-1}\beta_k|_{\hti_i=0}\label{hbf}\\
\chi^2_{min}&=&S_{f^2}-\tilde{F}_{ij}\hti_{min,i}\hti_{min,j}\label{x2bf},
\ea
where
\be
\beta_k|_{\hti_i=0}=-\alpha\left(S_{fz_k}+\sum_{n=1}^{n_{tot}}A_{nk}S_{f_n}\right)\label{bibf}
\ee
As can be seen from Eqs.~(\ref{fisherH}), (\ref{hbf}) and (\ref{bibf}) the various parameters scale differently with $\alpha$ or equivalently $h$, eg the the Fisher matrix scales as $\tilde{F}_{ij}\sim \alpha^2 \sim h^{-2}$, while the best fit parameters as $\hti_{min,j} \sim \alpha^{-1}\sim h$. On the other hand, as seen from Eq.~(\ref{x2bf}) the minimum chi square $\chi^2_{min}$ is invariant since the contributions from $\tilde{F}_{ij}$ and $\hti_{min,j}$ cancel out. This means that in this case the best-fit is degenerate with respect to $h$ and as a result we have to fix it to some value before the actual fit.

Finally, we can also rotate the parameters to a basis where they are not correlated with each other, as in Ref. \cite{Nesseris:2012cq}. To do so we define a new variable $s_i\equiv D_{ij}\left(\hti_j-\hti_{j, {\rm min}}\right)$, where $D_{ij}$ can be found by
decomposing the inverse Fisher matrix $\tilde{F}=\tilde{C}^{-1}=D^T D$ by
using Cholesky decomposition\footnote{Cholesky decomposition can easily be implemented
in computer programs such as Mathematica. For example, in the latter the Cholesky decomposition of a matrix $M=D^{T} D$ is given by $D=CholeskyDecomposition[M]$. This works both symbolically and numerically.}. Then, going to the new basis we have
\ba
s_i &\equiv& D_{ij}\left(\hti_j-\hti_{j, {\rm min}}\right)\label{transf1}\\
ds_1 ... ds_N&=&\left\vert D \right\vert d\hti_1 d\hti_2...d\hti_{n_{tot}} \\
\left\vert D \right\vert &=& \left\vert \tilde{F} \right\vert^{1/2}=\left\vert \tilde{C} \right\vert^{-1/2}. \label{transf3}
\ea
Also, it can be easily shown that for the uncorrelated parameters $s_i$ we have
\be
\chi^2=\chi^2_{min}+s_1^2+s_2^2+...+s_{n_{tot}}^2.
\ee
The Fisher matrix of the original $h_i$ parameters will be given by $F=J^T \tilde{F} J$,
where $\tilde{F}$ is given by Eq.~(\ref{fisherH}) and $J^i_j=\frac{\partial \hti^i}{\partial h^j}=-\hti_j^2\delta_{ij}$ is the Jacobian of the transformation.

In order to find the dark energy EOS parameter $w$ we can use Eqs.~(\ref{eq:rhoDE1}) and (\ref{eq:rhoDE2}). It is important to realize that the values of the $h_i=H_i/H_0$ parameters actually correspond to the average redshift in the bin, i.e. $z_{eff,i}=\frac{1}{2}\left(z_{i-1}+z_i\right)$, so that we can evaluate Eq.~(\ref{eq:rhoDE1}) at two different redshifts $z_{eff,i-1}$ and $z_{eff,i}$, and subtract to get
\be
\int_{z_{eff,i-1}}^{z_{eff,i}}\frac{1+w(z)}{1+z}\rmd z=\frac13 \ln\left(\frac{H_i^2/H_0^2-\omo (1+z_{eff,i})^3}{H_{i-1}^2/H_0^2-\omo (1+z_{eff,i-1})^3}\right).
\ee
In general we cannot evaluate the left hand side of the above equation,
but if we use the mean value theorem for integration, then we can write it as
\ba
\int_{z_{eff,i-1}}^{z_{eff,i}}\frac{1+w(z)}{1+z}\rmd z&=&\left[1+w(x)\right]\int_{z_{eff,i-1}}^{z_{eff,i}}
\frac{1}{1+z}\rmd z \nn\\
&=&\left[1+w(x)\right]\ln\left(\frac{1+z_{eff,i}}{1+z_{eff,i-1}}\right),
\ea
where $x\in(z_{eff,i-1},z_{eff,i})$.

For example, using an evolving DE equation of state like $w(z)=w_0+w_1 \frac{z}{1+z}$ it is easy to calculate $x$ using the above formulas. Using the fact that for equal sized bins we have $z_i=i dz$  and that $z_{eff,i}\equiv \frac{1}{2}\left(z_{i-1}+z_i\right)=(i-1/2)dz$ then, in this case we find that
\ba
x&\simeq& (-1 + i) dz - dz^2/6 +..., \nn\\
&\simeq& z_{i-1}- dz^2/6 +...,
\ea
where the first term is the redshift of the lower bin and the second is a correction.
Surprisingly, in this case $x$ does not depend on on the parameters $w_0$ and $w_1$.
For small bins or large redshifts, the last term usually is negligible and
we have confirmed this with numerical tests. For example, for $(z_1,z_2)=(0.1,0.2)$
the two terms are 0.1 and -0.0016 respectively, while for larger bins like $(z_1,z_2)=(0.1,0.5)$
the terms are 0.1 and -0.027, thus confirming our assumptions.

Of course, in the case of rapidly evolving equation of state these assumptions do not necessarily hold any more. This can easily be seen by considering a model of the form $w(z)=w_0+w_1 z +\frac12 w_2 z^2+...$, where the second derivative $w''(z=0)\equiv w_2$ is not necessarily small, i.e. we cannot assume $|w_2|\ll 1$. Then, the parameter $x$ is given by
\ba
x&\simeq& dz (i-1) + \frac{1}{24} dz^2 \left(\frac{w_2}{w_1}-4\right)+..., \nn \\
&\simeq& z_{i-1} + \frac{1}{24} dz^2 \left(\frac{w_2}{w_1}-4\right)+... .
\ea
Clearly, in this case there might be a small effect due to the cosmology. However, models with fast transitions of the equation of state seem to be disfavored by observations \cite{DeFelice:2012vd}, so in what follows we will assume that $w(z)$ may only be evolving slowly. Therefore, to excellent  approximation we consider that $x \sim z_{i-1}$ and the DE equation of state at a bin $n$ will be given by
\ba
w_n(x_i)&=&-1+\frac{\ln\left(\frac{H_i^2/H_0^2-\omo (1+z_{eff,i})^3}{H_{i-1}^2/H_0^2-
\omo (1+z_{eff,i-1})^3}\right)}{\ln\left(\frac{1+z_{eff,i}}{1+z_{eff,i-1}}\right)^3}, \nn \\
x_i &\simeq& z_{i-1}.
\ea
Intuitively the above result can be understood as follows. The parameters $H_i$ correspond to the redshift in the middle of the bins, so taking their differences produces a result that corresponds to the sides of the bins.

\subsection{The luminosity distance $d_L(z)$\label{app-PCAdl}}

As mentioned earlier in the paper, we do not bin directly the luminosity distance $d_{L(z)}$ but rather the distance moduli $\mu(z)$:
\be
\mu(z) = \sum_{i=1}^{n_{tot}}\mu_{i}\theta(z_i),
\ee
where $\mu_i$ are constant in each redshift bin $z_i$, $n_{tot}$ is the total number of bins and $\theta(z_i)$ is the theta function, i.e. $\theta(z_i) = 1$ for $z_{i-1}< z \leq z_{i}$ and $0$ elsewhere. In this case, the chi squared can be written as
\be
\chi^2=\sum_{i=1}^N\left(\frac{\mu_{obs,i}-\mu_i}{\sigma_{i}}\right)^2.
\ee
Clearly, the $\chi^2$ is linear with respect the parameters $\mu_n$, so in this case we can find closed-form analytical expressions for the best-fit parameters. First, we will make the following definitions
\ba
S&\equiv& \sum_{i=1}^N\frac{1}{\sigma_i^2} ~~~~~~~~~~~S_\mu \equiv \sum_{i=1}^N\frac{\mu_{obs,i}}{\sigma_i^2},\\
S_{\mu^2} &\equiv& \sum_{i=1}^N\frac{\mu_{obs,i}^2}{\sigma_i^2}~~~~~~~
S_n\equiv \sum_{j=1,\textrm{n bin}}\frac{1}{\sigma_j^2},\label{eq:SnApp}\\
S_{\mu_n} &\equiv& \sum_{j=1,\textrm{n bin}}\frac{\mu_{obs,j}}{\sigma_j^2},
\ea
where $\sum_{j=1,\textrm{n bin}}$ is meant to sum over only those points
in the $n^{\textrm{th}}$ bin, by which it follows that
\ba
\sum_{n=1}^{n_{tot}}S_n&=&S,\\
\sum_{n=1}^{n_{tot}}S_{\mu_n}&=&S_\mu\,,
\ea
where $n_{tot}$ is the total number of bins.
With these in mind we can now find the best fit by taking the derivatives
with respect to the parameters
\ba
\partial_{\mu_n}\chi^2|_{min}&=&\sum_{i=1}^N 2\left(\frac{\mu_{obs,i}-\mu_i}{\sigma_{\mu,i}^2}\right) \left(-\frac{\partial \mu_i}{\partial \mu_n}\right)\nn\\
&=& -2\sum_{j=1,\textrm{n bin}}\left(\frac{\mu_{obs,i}-\mu_n}{\sigma_{\mu,i}^2}\right) \\
&=& -2S_{\mu_n}+ 2\mu_n S_n\\
&=&0, \label{chi2deriv}
\ea
where we have used  the fact that $\frac{\partial \mu_i}{\partial \mu_n}=\delta_{i,n}$ and
that the first derivative should be zero at the minimum.
Then, Eq.~(\ref{chi2deriv}) can readily be solved to yield
\be
\mu_n=\frac{S_{\mu_n}}{S_n}. \label{bestmu}
\ee
A similar calculation reveals that the chi square and its value at the minimum are
\ba
\chi^2 (\mu_n)&=&S_{\mu^2}-2 \sum_{k=1,\textrm{n bin}}\mu_k S_{\mu_k}+\sum_{k=1,\textrm{n bin}}\mu_k^2 S_k \label{chi2dL}\\
\chi^2_{min}&=&S_{\mu^2}-\sum_{n=1}^{n_{tot}}\frac{S_{\mu_n}^2}{S_n}.
\ea
The errors on the best-fit parameters can be estimated by direct error propagation,
see chapter 15 of Ref.~\cite{press92}, as
\ba
\sigma_{\mu_n}^2&=&\sum_{i=1}^N\sigma_i^2\left(\frac{\partial \mu_n}{\partial \mu_{obs,i}}\right)^2\nn\\
&=&\sum_{i=1}^N \sigma_i^2 \left(\frac{1}{S_n}\frac{\partial S_{\mu_n}}{\partial \mu_{obs,i}}\right)^2 \nn\\
&=& \sum_{i=1}^N \sigma_i^2 \frac{1}{S_n^2} \left(\sum_{j=1,\textrm{n bin}}\frac{1}{\sigma_j^2}\frac{\partial \mu_{obs,j}}{\partial \mu_{obs,i}}\right)^2 \nn\\
&=& \sum_{i=1}^N \sigma_i^2 \frac{1}{S_n^2} \left(\sum_{j=1,\textrm{n bin}}\frac{1}{\sigma_j^2}\delta_{i,j (n bin)}\right)\left(\sum_{k=1,\textrm{n bin}}\frac{1}{\sigma_k^2}\delta_{i,k (n bin)}\right) \nn\\
&=& \frac{1}{S_n^2}\sum_{i=1}^N \sum_{j=1,\textrm{n bin}}\sum_{k=1,\textrm{n bin}}\sigma_i^2 \frac{1}{\sigma_j^2}\frac{1}{\sigma_k^2}\delta_{i,j (n bin)}\delta_{i,k (n bin)}\nn\\
&=&\frac{1}{S_n}.\label{muerrors}
\ea
We find that the results of Eqs.~(\ref{bestmu}) and (\ref{muerrors}) are in agreement with the ones found by considering the binning of data. Following a more direct approach, by calculating directly the Fisher and the covariance matrices from Eq.~(\ref{chi2dL}) evaluated at the minimum, we get
\ba
F_{nk}&=&\frac12 \partial_{nk}^2\chi^2|_{\textrm{min}}=\textrm{diag}\left(S_1,S_2,...,S_{n_{tot}}\right)\\
C_{nk}&=&F_{nk}^{-1}=\textrm{diag}\left(S_1^{-1},S_2^{-1},...,S_{n_{tot}}^{-1}\right),
\ea
where $\textrm{diag}\left(...\right)$ is a $n_{tot} \times n_{tot}$ diagonal matrix.
The diagonal terms of the covariance matrix are the errors $\sigma_{\mu_n}^2$ and they are in exact agreement with Eq.~(\ref{muerrors}).
Finally, we should note that the covariance matrix is diagonal, which means that the parameters are already uncorrelated and we do not have to follow the PCA approach in this case.

In order to extract the cosmology we can invert the equation of the distance modulus and
we find the dimensionless luminosity distance to be
\be
D_{L,i}=10^{\frac{\mu_i-\mu_0}{5}}\label{eq:DLi}
\ee
where $\mu_0\simeq42.384-5\log_{10}h$.

\section{The chi square for the reconstructions}
Another way to test which of the four methods reconstructs the ``real" cosmologies the best,
is to use a chi square:
\ba
\chi^2&=&\sum_{i=1}^{n_{tot}}\frac{(y(z_i)-y_{real}(z_i))^2}{\sigma(y(z_i))^2}
\ea
where $y(z_i)$ are the reconstructed (best-fit or derived) parameters in each
bin i.e. $w,\,q,\,H,\,d_L$, $y_{real}(z_i)$ are the ``real" values of the parameters,
and $n_{tot}$ is the number of points. The results for the $\chi^2$ for the three cosmologies are shown in Tables \ref{tab:chi2-LCDM}, \ref{tab:chi2-CPL} and \ref{tab:chi2-f_R} respectively. The columns indicate the methods use to fit the data, while the row the reconstructed parameters. However, this method suffers from several problems.
For example, let us consider the results for the \lcdm model, shown in Table~\ref{tab:chi2-LCDM}; as it can be seen the worst results are given when we try to reconstruct the luminosity distance, for which we have a $\chi^2$ of about $4000$. How could be this possible? The reason is that the errors of the binned luminosity distance $d_L$ are very small, of about $10^{-3}$ and the $\chi^2$ is proportional to the inverse of the square of the
errors; however this is not the only reason why the $\chi^2$ is extremely large.

As we can see from the fourth row in Fig.~\ref{fig:pca}, the best fit values of the
luminosity distance are far from the theoretical curve (red dashed line in the same figure).
It is worth mentioning that here we are trying to reconstruct functions using mock catalogs
evaluated with a specific cosmology, with the hope of getting the initial cosmology at the end, in other words verifying that our reconstruction methods indeed work as advertised. The difference from the theoretical curve and the PCA values make the $\chi^2$ explodes as the numerator of the $\chi^2$ will not be sufficiently small to kill the $\sigma_i^2$ in the denominator. This is the opposite to what it is usually done when we deal with data, where we have a dataset and we try to find the best fit, which mean to find those curves that better describe the data within the errors. In the latter case the $\chi^2$ will be in general small
even if the errors are extremely small.

In this work we found that the general $\chi^2$ fails to describe the goodness or quality of our analysis, while the risk seems to be a more suitable parameter. As a final remark, we notice that the values shown in Table~\ref{tab:chi2-LCDM} follow a general trend which is the opposite to the risk, i.e., when we go to a more complicated function then the errors increase and the $\chi^2$ decreases (fourth column in Table~\ref{tab:chi2-LCDM} from down to up). This effect can also be explained, as it was mentioned in the previous paragraph, by the fact that the very small or large errors affect the estimation of the $\chi^2$ giving artificially large or small values even when the fit is obviously quite good.

%%%%%%%%%%%%%%%%
\begin{table}[t!]
\begin{centering}\begin{tabular}{|c|c|c|c|c|}
\hline
\multicolumn{5}{|c|}{$\chi^2$ for $\Lambda$CDM}  \tabularnewline
\hline
\backslashbox{{\small Derived param.}}{Piecewise} & $w$ & $q$ & $H$ & $D_L$  \tabularnewline
\hline
$w$ & $20.991$ & $1.126$ & $6.248$ & $29.034$  \tabularnewline
\hline
$q$ & $1.966$ & $15.693$ & $5.120$ & $40.129$  \tabularnewline
\hline
$H$  & $3.373$ & $1.981$ & $32.046$ & $471.493$ \tabularnewline
\hline
$D_L$  & $8.067$ & $2.672$ & $691.052$ & $3959.100$ \tabularnewline
\hline
\end{tabular}\par\end{centering}
\caption{The values of the $\chi^2$ for the $\Lambda$CDM cosmology. The rows indicate the derived parameters, while the columns indicate the piecewise method.
\label{tab:chi2-LCDM}}
%\end{table}
%%%%%%%%%%%%%%%%%%%
%
%%%%%%%%%%%%%%%%%
%\begin{table}
\begin{centering}\begin{tabular}{|c|c|c|c|c|}
\hline
\multicolumn{5}{|c|}{$\chi^2$ for CPL}  \tabularnewline
\hline
 \backslashbox{{\small Derived param.}}{Piecewise}& $w$ & $q$ & $H$ & $D_L$  \tabularnewline
\hline
$w$ & $22.908$ & $1.391$ & $5.615$ & $26.110$  \tabularnewline
\hline
$q$ & $3.662$ & $22.386$ & $3.253$ & $39.005$  \tabularnewline
\hline
$H$  & $4.954$ & $2.569$ & $30.992$ & $398.359$ \tabularnewline
\hline
$D_L$  & $7.141$ & $3.457$ & $674.500$ & $3967.190$ \tabularnewline
\hline
\end{tabular}\par\end{centering}
\caption{The values of the $\chi^2$ for the CPL cosmology. The rows indicate the derived parameters, while the columns indicate the piecewise method.
\label{tab:chi2-CPL}}
%\end{table}
%%%%%%%%%%%%%%%%%%%
%%%%%%%%%%%%%%%%%
%\begin{table}
\begin{centering}\begin{tabular}{|c|c|c|c|c|}
\hline
\multicolumn{5}{|c|}{$\chi^2$ for $f(R)$}  \tabularnewline
\hline
 \backslashbox{{\small Derived param.}}{Piecewise}& $w$ & $q$ & $H$ & $D_L$  \tabularnewline
\hline
$w$ & $9.081$ & $0.991$ & $6.199$ & $29.391$  \tabularnewline
\hline
$q$ & $1.951$ & $6.726$ & $5.405$ & $39.925$  \tabularnewline
\hline
$H$  & $2.400$ & $1.138$ & $32.009$ & $471.299$ \tabularnewline
\hline
$D_L$  & $8.059$ & $0.692$ & $687.274$ & $3958.900$ \tabularnewline
\hline
\end{tabular}\par\end{centering}
\caption{The values of the $\chi^2$ for the $f(R)$ cosmology. The rows indicate the derived parameters, while the columns indicate the piecewise method.
\label{tab:chi2-f_R}}
\end{table}
%%%%%%%%%%%%%%%%%%

{}

\end{document}